\documentclass[reprint,amsmath,amssymb,aps,superscriptaddress]{revtex4-2}

\usepackage{graphicx}
\usepackage{dcolumn}
\usepackage{amsmath}
\usepackage{bm}
\usepackage{mathrsfs}
\usepackage{txfonts}
\usepackage{epstopdf}
\usepackage{subfigure}
\usepackage{caption}
\usepackage{float}

\begin{document}
\title{Role of line defect in the bandgap and transport properties of silicene nanoribbons}
	
\author{Fei Wan}
	
\author{Xinru Wang}
	
\author{Yawen Guo}
	
\author{Jiayan Zhang}
	
\author{ZhengCheng Wen}
\email[]{wenzc@hdu.edu.cn}

\author{Yuan Li}
\email[]{liyuan@hdu.edu.cn}
\affiliation{College of science, Hangzhou Dianzi University, Hangzhou, Zhejiang 310018, China}
	
\date{\today}
	
\begin{abstract}
By using the tight-binding model and non-equilibrium Green's function method (NEGF), we study the band structures and transport properties of a silicene nanoribbon with a line defect where a bulk energy gap is opened due to the sublattice symmetry breaking.
The flat subband bends downwards or upwards due to the effect of the line defect. The spin-orbit coupling induces quantum spin Hall states. Especially, the energy band depends on the distance between the line defect and the edge of the nanoribbon. The effects of the on-site energies on the band spectra of the two defect configurations are different. There always exists one band gap for different on-site energies for the defect configuration of case 1. However, a gapless state and a band gap can be modulated by changing the on-site energy, the sublattice potential and spin-orbit couplings for the defect configuration of case 2. Accordingly, the variation trends of the conductance including zero conductance can be well understood in terms of the combined effect of the sublattice potential, the on-site energy and spin-orbit couplings on the band structures. Thus it is easy and effective to modulate the transport property of the silicene nanoribbon with the defect configuration of case 2 by utilizing the sublattice potential, the on-site energy and spin-orbit couplings. This study is of great significance for the fabrication and the modulation of the transport property of silicene-based devices.

\end{abstract}
	
	
\maketitle
	
\section{Introduction}
Silicene is a low-buckle honeycomb structure formed from a monolayer silicon atoms. In recent years, after being synthesized on metal surfaces successfully~\cite{Aufray2010,Padova2010,Boubekeur2010}, it has attracted much attention between researchers of both theoretical~\cite{Fagan2000,Cahangirov2009} and experimental fields~\cite{Chen2012,Guzm2007}.Its low-buckled geometry creates a relatively large gap opened by the spin-orbit coupling at Dirac points~\cite{Liu2011,Drummond2012}. It is also reported that the size of band gap increases as the amplitude of the electric field increases and a phase transition from a topological insulator to a band insulator will happen in the process~\cite{Xing-Tao2013,Ezawa_2012}. What's more, silicene stimulates the development of many fields involved with valley-polarized quantum anomalous Hall effect~\cite{Ezawa_2012}, quantum spin Hall effect~\cite{Liu2011a,Tabert2013}, spin and valley polarization~\cite{Ezawa2012a,Missault2015}, etc.
	
Recently, the extended line defects (ELDs) in silicene have been extensively investigated according to first-principles calculations~\cite{Ghosh2015}, and the 5-5-8 ELD (abbreviated as "line defect" in the following) was found to be the most stable and most readily formed structure. The spin and valley polarization of the silicene with line defects have been investigated theoretically~\cite{Ren2_2018,YANG2015396,wang22018}. The formation of a line defect can be visualized as the stitching of the zigzag edges of two Si grains adsorbed by Si atoms. the side of the line defect shows pseudo-edge-state-like behavior and the grain boundaries of the zigzag edge act as the pseudo-edge~\cite{Ghosh2015}.These results show that the silicene with a line defect has a great application prospect. Therefore, motivated by the idea of finding a potential application for a line defect in silicene, we extend the position-dependent effect and the investigation of line defects in silicene nanoribbons with sublattice symmetry breaking.
	
In this work, we use the tight-binding model and NEGF~\cite{1996Datta} method to study how a line defect affects the transport property of a silicene nanoribbon with zigzag edges in two configurations of sublattice symmetry breaking. One configuration has a normal silicene nanoribbon with a line defect, another case has mirror symmetry~\cite{2019Controllable} . This paper describes some novel states arising from the band gap induced by line defects, and these states can be effectively modulated by adjusting the distance between the line defect and the zigzag edges,
the gate voltage embedded in the line defect, and the parameter of SOC in silicene nanoribbons. Based on these results, a device model is proposed at the end of this paper.
	
\section{THEORETICAL MODEL AND FORMULA}
\begin{figure}[t]
	\centering
	\includegraphics[scale=0.45,trim=2 0 0 2,clip]{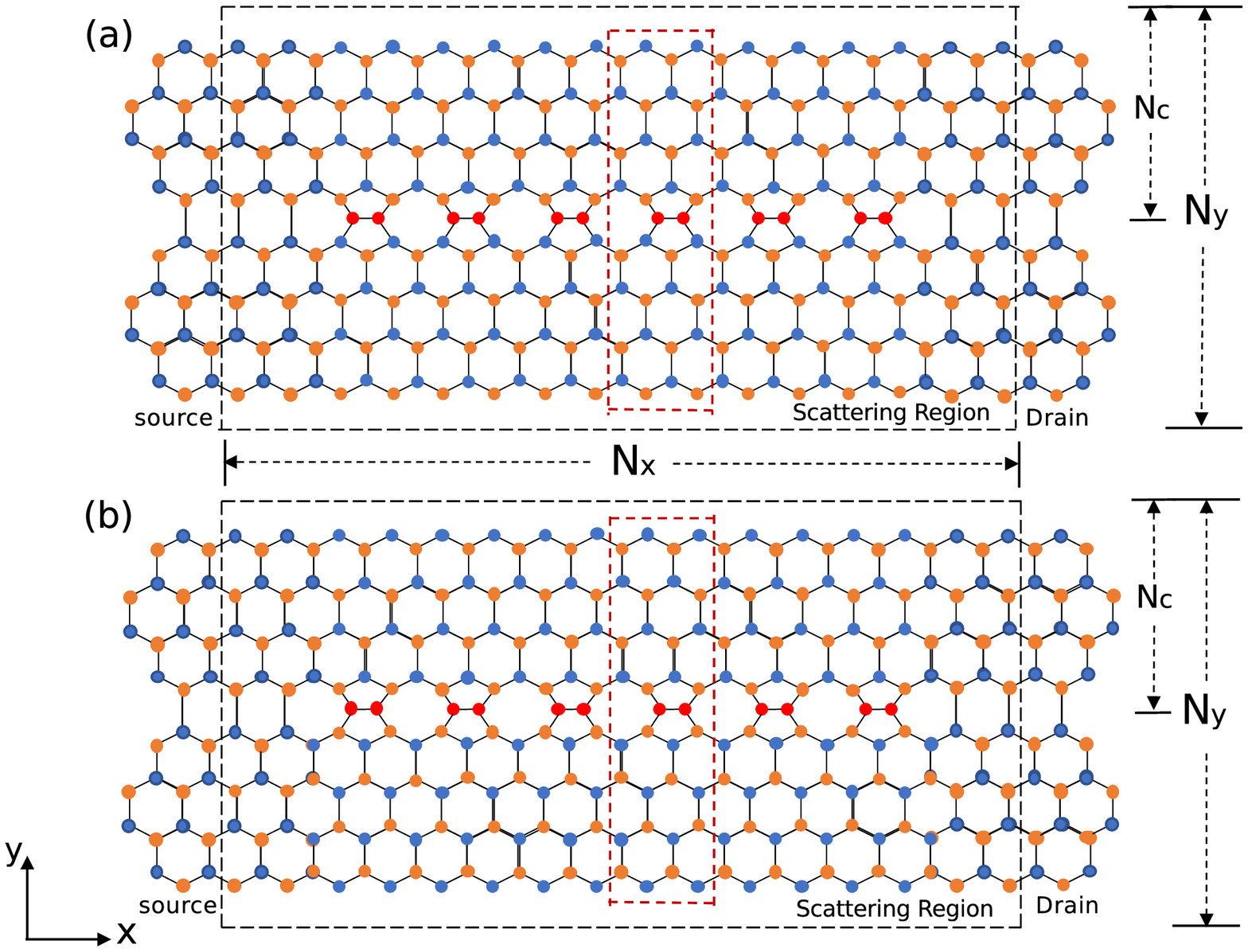}%

	\includegraphics[scale=0.5]{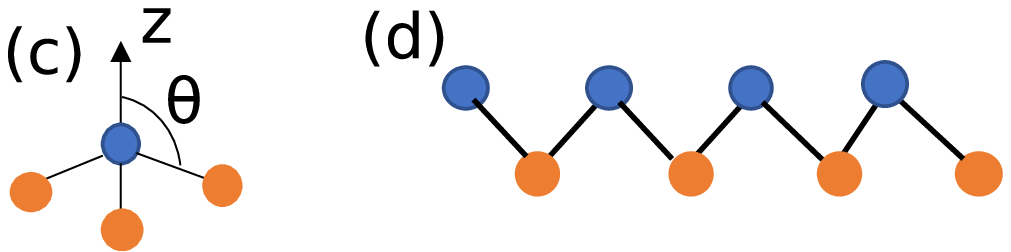}%
	\caption{A schematic diagram for an infinitely long
		zigzag silicene nanoribbon with a line defect (red). There are two cases:
		(a) the line defect connects A (orange) and B (blue) sublattice sites;
		(b) the line defect connects two of the same type of sublattice
		sites (A-A). (c) $\theta$ is defined as the angle between Si-Si bond and z axis is normal to the plane.
 (d) The side view of the low-buckled silicene structure.}
\label{fig:structure}
	\end{figure}

By using the four-band second-nearest-neighbour tightbinding model~\cite{liyuan2018}, as shown in Fig.~\ref{fig:structure}, the Hamiltonian can be written as the following form\\
\begin{eqnarray}
H&=&-t\sum_{\langle i,j\rangle\alpha}c_{i\alpha}^{\dag}c_{j\alpha}-i\frac{2}{3}t_{R2}\sum_{\langle\langle i,j\rangle\rangle\alpha\beta}\mu_{i}
c_{i\alpha}^{\dag}\left(\vec{\sigma}\times\hat{d}_{ij}\right)_{\alpha\beta}^zc_{i\beta}\nonumber\\
&&+i\frac{t_{so}}{3\sqrt{3}}\sum_{\langle\langle i,j\rangle\rangle\alpha\beta}\nu_{ij}c_{i\alpha}^{\dag}\sigma_{\alpha\beta}^{z}c_{i\beta}
-\sum_{i\alpha}\mu_{i}a_{z}E_{z}c_{i\alpha}^{\dag}c_{i\alpha},
\end{eqnarray}
where $c_{i\alpha}^{\dag}~(c_{i\alpha})$ refers to the creation (annihilation) operator with spin index $\alpha$ at site $i$, and $\langle i,j\rangle/\langle\langle i,j\rangle\rangle$ run over all the nearest- or next-nearest-neighbor hopping sites. The first term is the nearest-neighbor hopping with the transfer energy $t$. The second term represents the Rashba spin-orbit coupling with $\mu_{i}=\pm1$ for the A (B) site, where $\hat{d}_{ij} = \hat{\mathbf{d}}_{ij}/|\hat{\mathbf{d}}_{ij}|$ refers to the unit vector connecting the two next-nearest-neighboring sites. The third term denotes the effective spin-orbit coupling with the hopping parameter $t_{so}$, where $\sigma=\left(\sigma_{x},\sigma_{y},\sigma_{z}\right)$ are the spin Pauli matrix operators, and $\nu_{ij}=\pm1$ for the anticlockwise (clockwise) hopping between the next-nearest-neighboring sites with respect to the positive $z$ axis. The fourth term describes the contribution of the staggered sublattice potential due to the perpendicular electric field~\cite{Missault2015,Gated}, with $a_{z} = 0.23 \mathrm{{\AA}}$ being the distance of the two sublattice planes. The values of the hopping parameters are $t=1.09 \mathrm{eV}$, $t_{so}=3.9 \mathrm{meV}$, and $t_{R2}=0.7 \mathrm{meV}$. A theoretical investigation\cite{Ghosh2015} has shown that the two nearest Si atoms in the defect region are relatively identical to those in the pristine region and that all Si atoms remain in sp$^{2}$-sp$^{3}$ hybridized states. Therefore, the nearest-neighbor-hopping term $t$ can be nearly identical to that of defect-free silicene systems.
	
For this Hamiltonian, we consider a general situation in which the bulk lattice can be subject to a staggered sublattice potential\cite{Song2012a}: $V_{i} =\Omega$ for lattice sites A, $V_{i} =-\Omega$ for lattice sites B, and $V_{i} = E_{d}$ for the line defect sites ( the red sites). Here,we use $\Omega=a_{z}E_{z}$ sketched in Fig. 1. For simplicity, in the following description of this paper, sublattice A (B) is introduced to represent the blue (orange) lattice sites. Note that sublattice sites A (or B) can possess either positive or negative staggered potentials by changing the sign of the parameter $\Omega$.
	
The conductance is calculated by using the Landauer-B\"{u}ttiker formalism,
\begin{eqnarray}
	G_{LR}=\frac{e^{2}}{h}\mathrm{T}=\frac{e^{2}}{h}\mathrm{Tr}[\Gamma_{L}G^{r}\Gamma_{R}G^{a}]
\end{eqnarray}
where $\mathrm{T}=\mathrm{Tr}[\Gamma_{L}G^{r}\Gamma_{R}G^{a}]$ is the transmission coefficient for electrons injected from the left lead (source) to the right lead (drain). The line-width functions $\Gamma_{L}/\Gamma_{R}= i(\Sigma_{L/R}^{r}-\Sigma_{L/R}^{a})$ are relative with the retarded/advanced self-energy $\Sigma_{L/R}^{r/a}$, which can be calculated numerically by solving the surface Green's function of the leads~\cite{Sancho}. Then we can calculate the
retarded and advanced Green's functions in terms of their definitions
\begin{eqnarray}
G^r(E)=[G^a(E)]^\dag=[E-H_c-\Sigma_L^r-\Sigma_R^r]^{-1},
\end{eqnarray}
where $H_c$ is the Hamiltonian of the central scattering region.
\begin{figure}[b]
\centering
\includegraphics[scale=0.39,trim=0 0 0 0,clip]{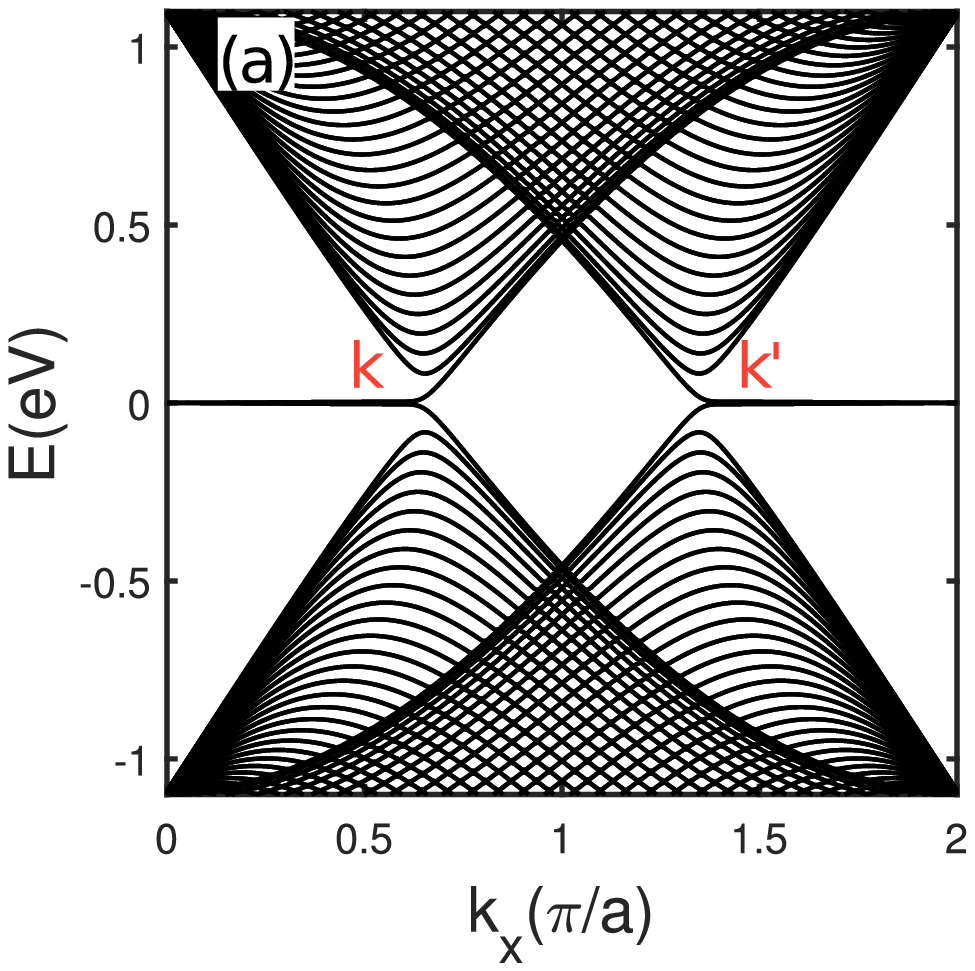}
\includegraphics[scale=0.39]{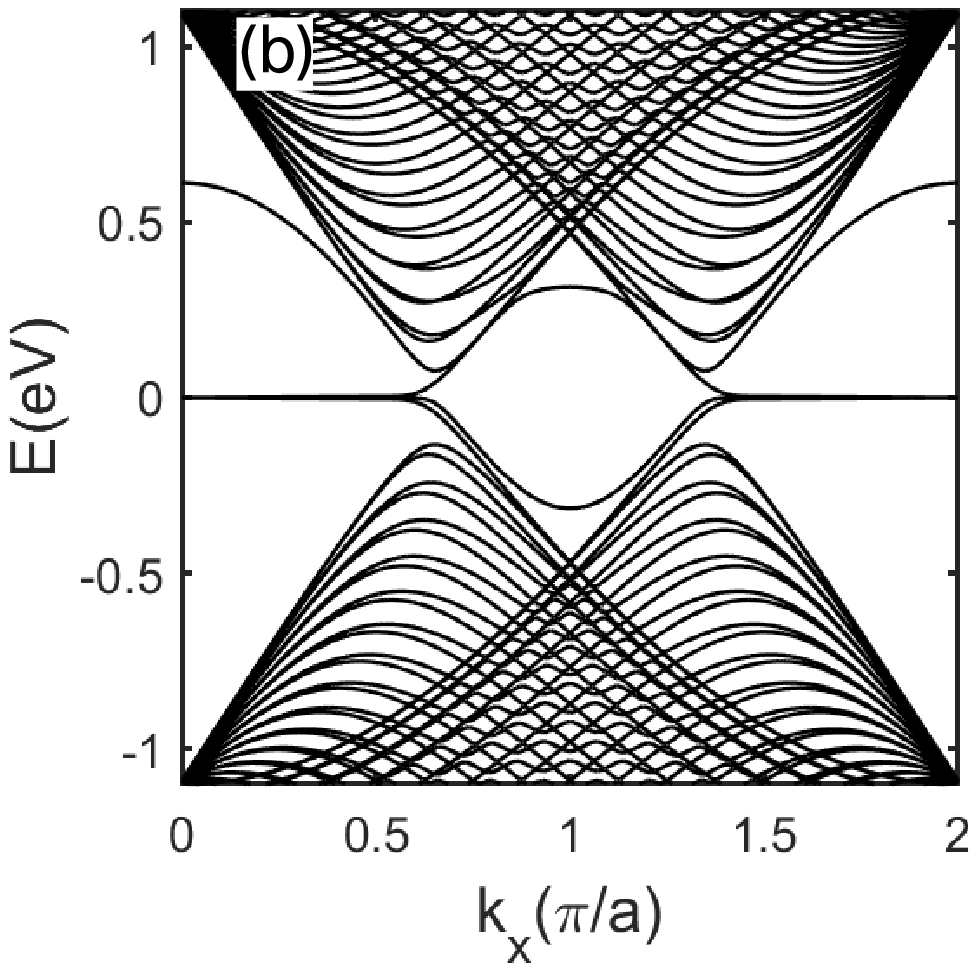}%
		
\includegraphics[scale=0.39]{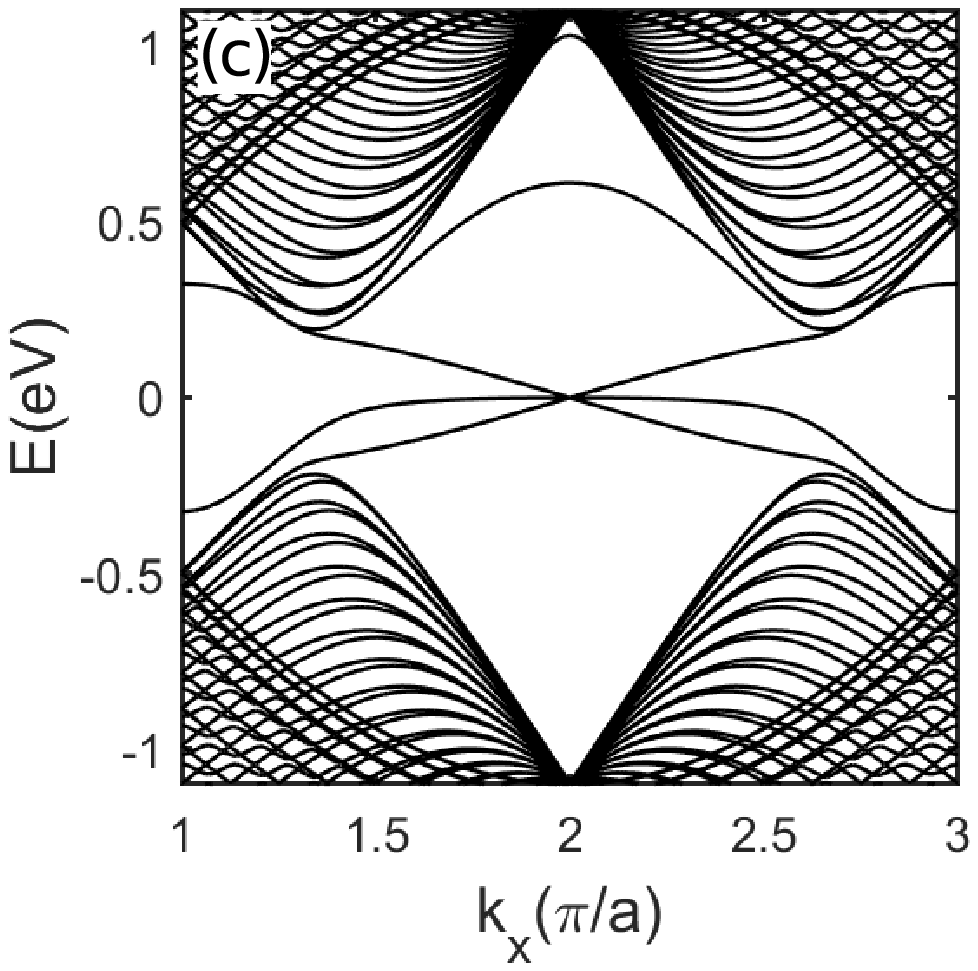}%
\includegraphics[scale=0.39]{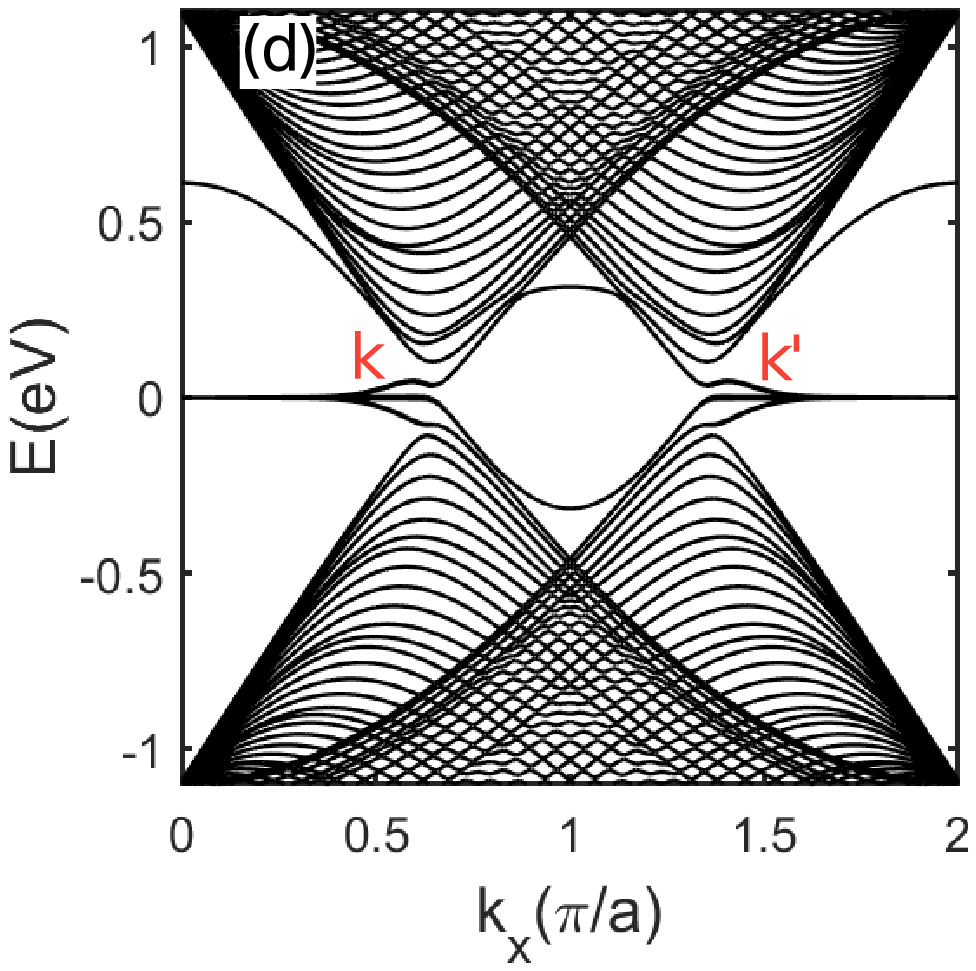}%
\caption{(a) Band structure of a zigzag nanoribbon for the case (a) without defects and $t_{so}=0$, (b) with a line defect at $N_{c}=8.69\mathrm{nm}$ and $t_{so}=0$,(c) a line defect at $N_{c}=8.69~\mathrm{nm}$ and $t_{so}=180 \mathrm{meV}$, (d) with a shifting defect at $N_{c}=4.35\mathrm{nm}$ and $t_{so}=0$. The on-site energy of the line defect is $ E_{d}=0$, $\Omega=0$, the length and width of the nanoribbon are chosen as $N_{x} = 121.80\mathrm{nm}$ and $N_{y} = 17.39\mathrm{nm}$.}
\label{fig:band1}
	\end{figure}
	
\section{NUMERICAL RESULTS AND DISCUSSION}
From the above tight-binding model for the situation shown in Fig.~\ref{fig:structure}, it is easy to get a nanoribbon geometry with zigzag edges. The length and width of the nanoribbon are denoted by $N_{x}$ and $N_{y}$, respectively. The distance  between the line defect and the nanoribbon edge denotes by $N_{c}$. The length $N_{x} = [(W_{x}-1)/2]\times\sqrt{3}a_{0}$ , the width $N_{y,c} \approx [(W_{y,c}-1)/2] \times3a_{0}$. $W_{x}$ and $W_{y}$ are the number of sites along $x$ and $y$ direction of the nanoribbon, respectively. $W_{c}$ is the number of sites between the line defect and the nanoribbon edge. Here, $W_{x}=621$, $W_{y}=53$, $W_{c}=27$, $a = 2\times\sqrt{3}a_{0}$ with $a_{0} = 0.223 \mathrm{nm}$ being the distance of two nearest silicon atoms in the pristine region. As seen in Fig.~\ref{fig:structure}(a), the line defect is adjacent to two different sublattice sites (A and B). In Fig.~\ref{fig:structure}(b), the line defect connects two identical-type sublattice sites (either A-A or B-B). For simplicity, in the following discussion, we regard the structure depicted in Fig.~\ref{fig:structure}(a) as case 1, and that in Fig.~\ref{fig:structure}(b) as case 2.

\subsection{Band spectra for different defect positions}
In Fig.~\ref{fig:band1}, we plot the band spectra of the two configurations
shown in Fig.~\ref{fig:structure} by changing the parameters $N_{c}$ and $t_{so} $ for $\Omega=0$. In Fig.~\ref{fig:band1}(a), one can see a flat band~\cite{YaoWang2009} at zero energy corresponding to edge states for the system without defect and $t_{so}=0$. When there exists a line defect, the particle-hole symmetry\cite{Bahamon2011,2011Electronic,2011Tunable,Pereira2006a} is broken in a new context. The flat band bends downward or upwards (see Fig.~\ref{fig:band1}(b)) when the line defect locates at the middle position of the nanoribbon, namely $N_{c} =8.69~\mathrm{nm}$. When the spin-orbit coupling is not zero, the flat bands gradually disappear and evolve into the quantum spin Hall (QSH) states~\cite{Liu2011a} with the parameters $N_{c} =8.69~\mathrm{nm}$ and $t_{so}=180~\mathrm{meV}$ [see Fig.~\ref{fig:band1}(c)]. One can see that the degeneracy of states at zero energy can be broken at $K$ and $K^{'}$ points by simply putting the defect closer to one edge, the case shown in Fig.~\ref{fig:band1}(d) when $t_{so}=0$ and the line defect is located at $N_{c}=4.35~\mathrm{nm}$, roughly halfway toward an edge from the middle in the same nanoribbon of $N_{y}=17.39\mathrm{nm}$.
Comparing with the results of Figs.~\ref{fig:band1}(b) and (d), the closer the line defect is to the edge, the greater the warpage of the energy band is. But the variation trends of the energy band qualitatively keep similar for different distances between the line defect and the edge. Especially, it is seen that each of subband splits into two subbands for the nanoribbon with defects, which is useful for us to understand the transport property of the systems.
	
\begin{figure}[t]
	\centering
	\includegraphics[scale=0.38]{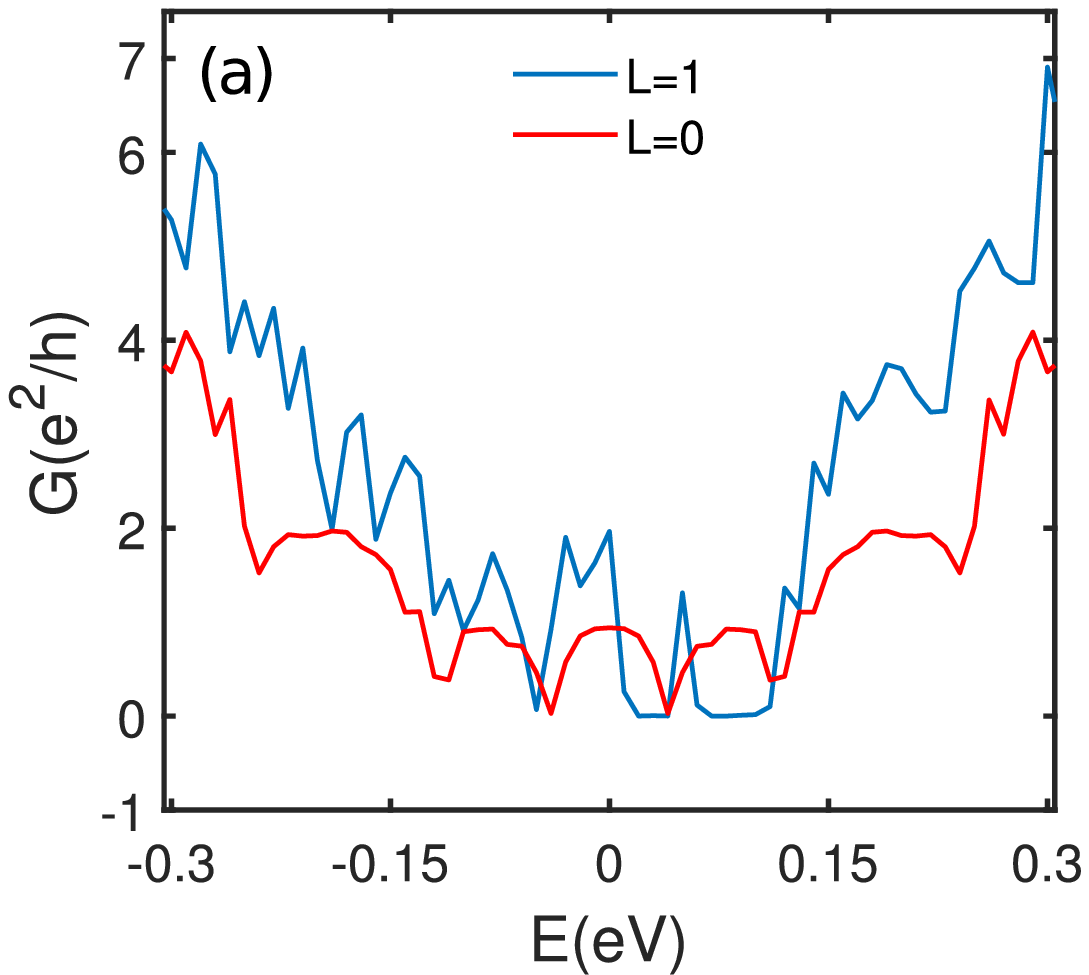}%
	\includegraphics[scale=0.38]{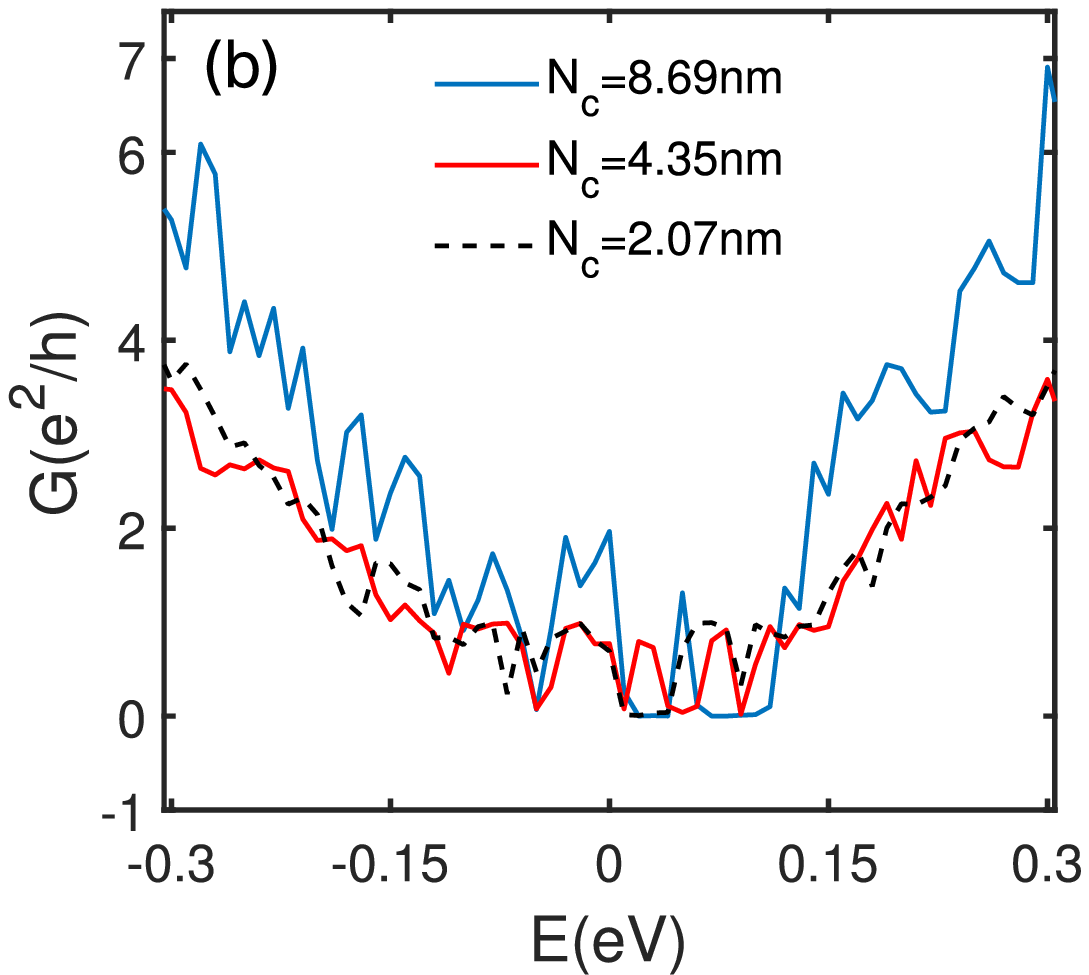}%
	\caption{The conductance $G$ is plotted as a function of Fermi energy for the case (a) with ($L=1$)/without ($L=0$) line defect, (b) with different positions of the line defect. The other parameters are the same as those in Fig.~\ref{fig:band1}.}
\label{fig:trans1}
\end{figure}
	
Next we study the position-dependent effect of the line defect on the conductance. In Fig.~\ref{fig:trans1}, we plot the conductance as a function of Fermi energy for different positions of the line defect. In the absence of the line defect ($L=0$), the conductance is symmetrical about the Fermi energy $E=0$. The conductance has a dip
near zero energy and then increases with increasing of Fermi energy. When a line defect occurs at middle position of the nanoribbon, the value of conductance becomes larger except two zero conductance regime due to the mismatch of the energy bands between leads and central scattering region [see blue curve in Fig.~\ref{fig:trans1}(a)]. It means that the line defect locating at the middle position can induce a larger transmission of electrons. When the line defect is shifted towards the edge[see red curve in Fig.~\ref{fig:trans1}(b)], for example $N_c=4.35 \mathrm{nm}$ and $N_c=2.07 \mathrm{nm}$, the conductance decreases and is asymmetrical about $E=0$ associating with the warpage of the energy band shown in Fig.~\ref{fig:band1}(d). Obviously, the conductance has a maximum value when the line defect is located at the middle position of the silicene nanoribbon. Thus, in the next part, we study the transport property of the silicene system by considering the middle configuration of the line defect.
	
\subsection{Band spectra for different on-site energies of the line defect}
\begin{figure}[t]
	\centering
	\includegraphics[scale=0.27]{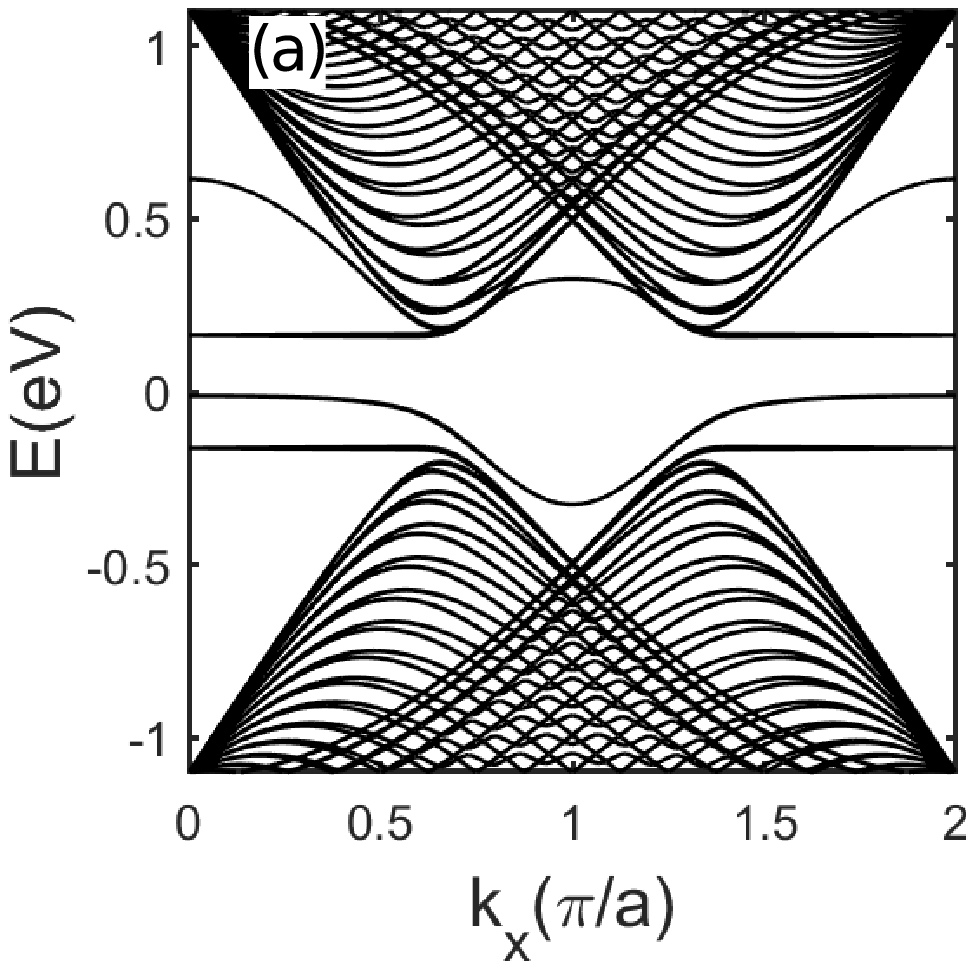}%
	\includegraphics[scale=0.27]{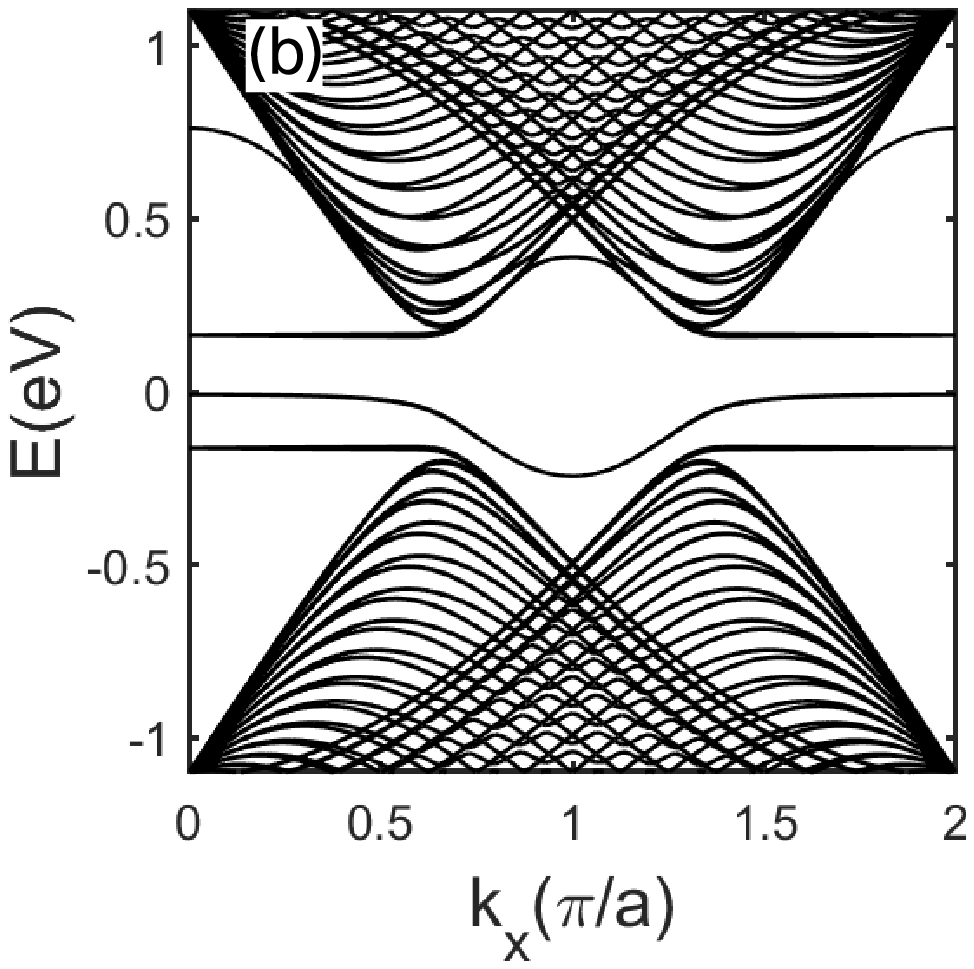}%
	\includegraphics[scale=0.27]{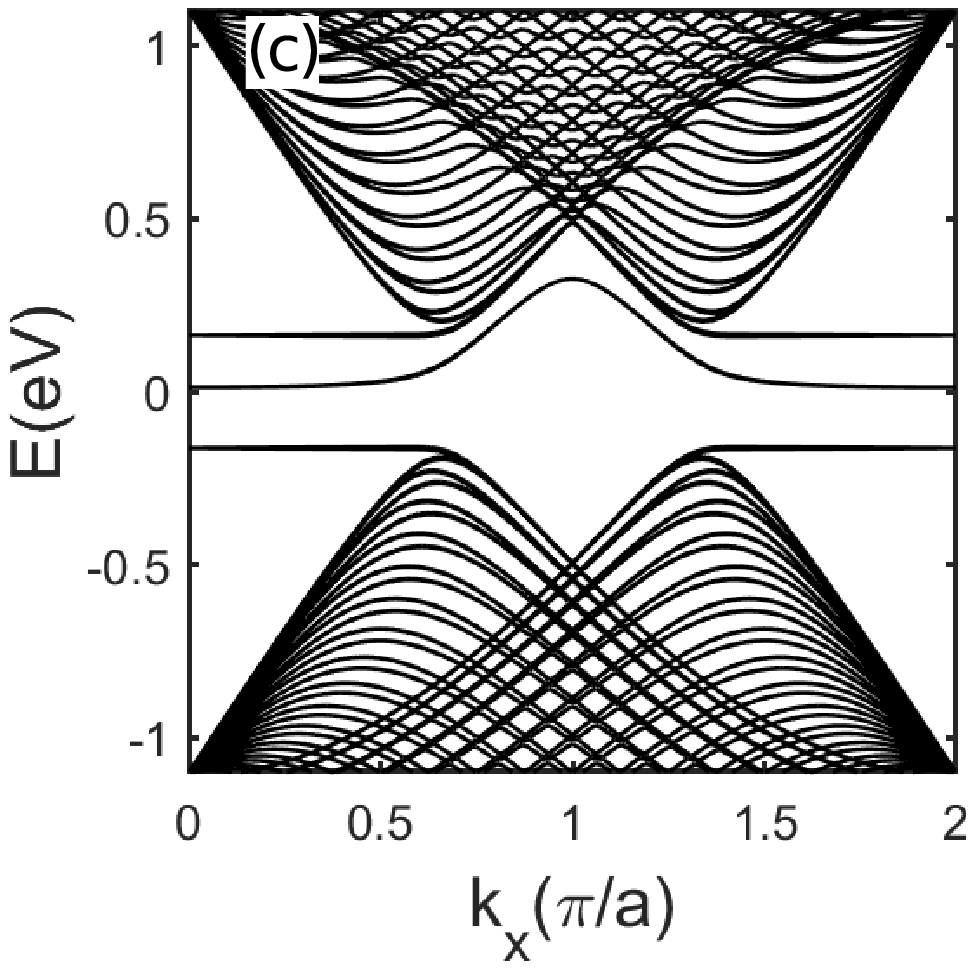}%

	\includegraphics[scale=0.27]{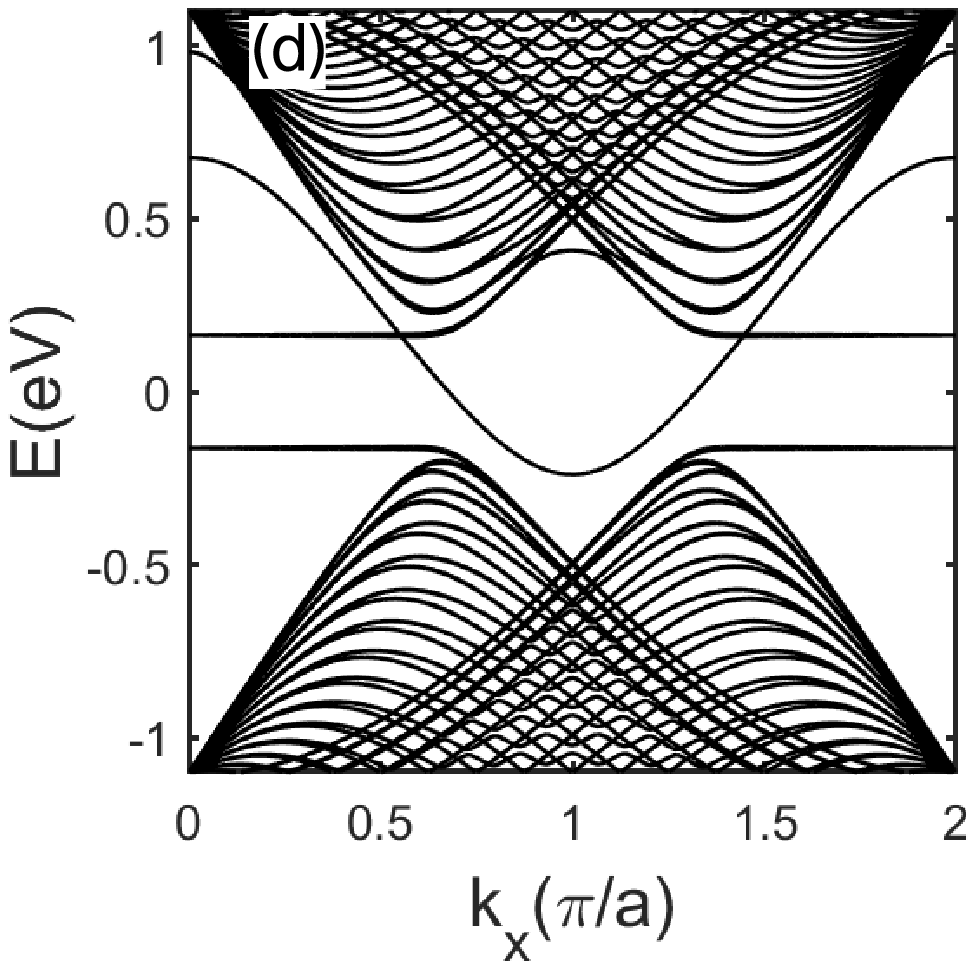}%
	\includegraphics[scale=0.27]{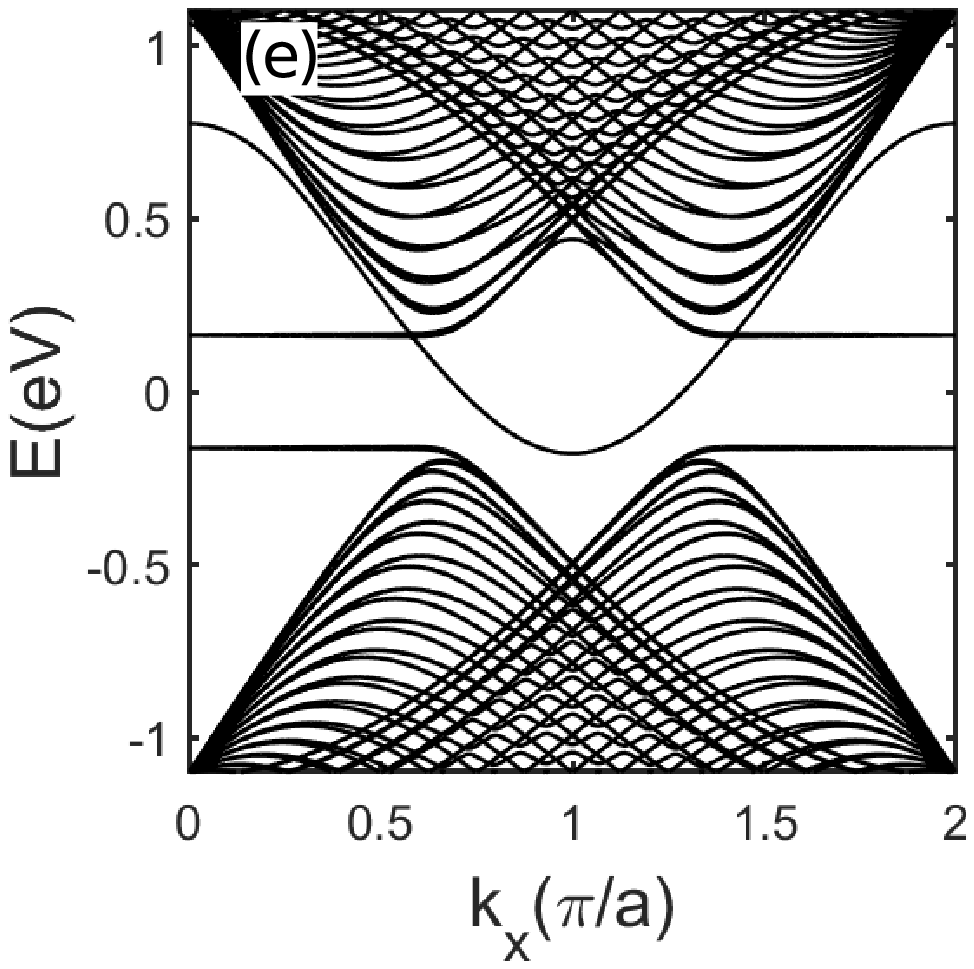}%
	\includegraphics[scale=0.27]{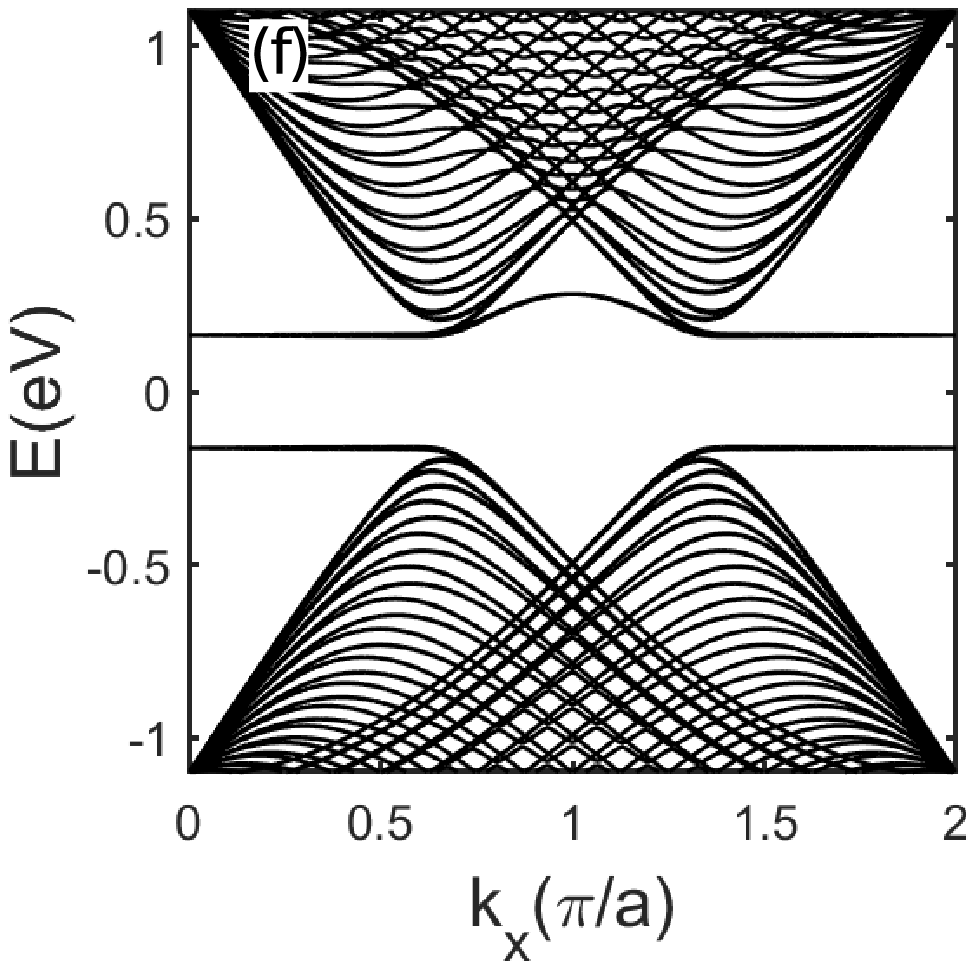}%
	\caption{Band spectra are plotted as a function of the wave vector $k_x$ for
	different on-site energies $E_{d}$ of the line defect. (a)-(c) associates with case 1 with $E_{d}=0,~0.3t$ and $3t$, respectively. (d)-(f) associates with case 2 with $E_{d}=0,~0.3t$ and $3t$. The other parameters are chosen as $\Omega=0.15t$, $t_{so}=3.9\mathrm{meV}$, $N_{x}=121.80\mathrm{nm}$ and $N_{y}=17.39\mathrm{nm}$.}
\label{fig:band2}
	\end{figure}

We further consider the effect of the middle line defect on the band structures. 	
In Figs.~\ref{fig:band2}(a)-(f), a bulk band gap can be observed for two cases by choosing a nonzero staggered potential. For case 1, when $\Omega=0.15t$, an extra flat band bends downward due to the line defect and fills approximately in half of the band gap from $E=-0.17\mathrm{eV}$ to 0 in Fig.~\ref{fig:band2}(a). However, when we change the on-site energy of the line defect to be $E_{d}=0.3t$ in Fig.~\ref{fig:band2}(b), the flat band basically keeps unchanged. But when the on-site energy increases to $E_{d}=3t$, the flat band bends upward in gap and fills in half of the band gap from 0 to $E=0.17\mathrm{eV}$ in Fig.~\ref{fig:band2}(c).
When the defect configuration is changed to case 2 with $ \Omega=0.15t$ as shown in Fig.~\ref{fig:band2}(d), an new quantum state appears again, but this state fully fills in the band gap. Thus in this case, the silicene nanoribbon becomes a gapless system. Some questions from these results naturally arise: What characteristics does the special state due to the line defect
shown in Fig.~\ref{fig:band2}(d) have, and are there some novel applications in silicene electronics related to these different band spectra? These are the interesting questions in this paper. We will focus mainly on the band spectrum in Fig.~\ref{fig:band2}(d) because the gapless state has, generally, much more interesting physics and all band spectra should have the same origin.
	
Focusing on the band structure in Fig.~\ref{fig:band2}(d), we can see that this special subband straddles the conduction band and the valence band. According to the band theory, the electrons occupying this subband can easily jump from the valence band to the conduction band and form free electrons. This reminds us of the nature of metals, where electrons in the valence band can easily jump to the conduction band to form free electrons. And this special subband has similar properties with metal. Due to the conductance has maximum values when the line defect locates at the middle position of the nanoribbon and the system shows metal properties. We can make silicene-based devices with large conductance by utilizing the novel gapless quantum state.
	
A basic question arises: can electric channels be realized by using line defects? That is to say, is there a simple way to open and close electric channels. Based on this idea, we study the band behavior of the line defect in silicene nanoribbons when the on-site energy of the line defect is tuned. In this way, the on-site energies can be continuously adjusted by the gate voltage embedded under the line defect.
As seen in Fig.~\ref{fig:band2}(e), when the on-site energy is changed to be $E_{d}=0.3t$, the gapless state does not disappear and still straddles the conduction band and the valence band. When $E_{d}=3t$, the gapless state disappears and a band gap is opened, which arises from the effect of the line defect and the electric field. Accordingly, the system has a transition from gapless state to the band insulator. What's more, the bending subband gradually becomes flat.
\begin{figure}[t]
	\centering
	\includegraphics[scale=0.27]{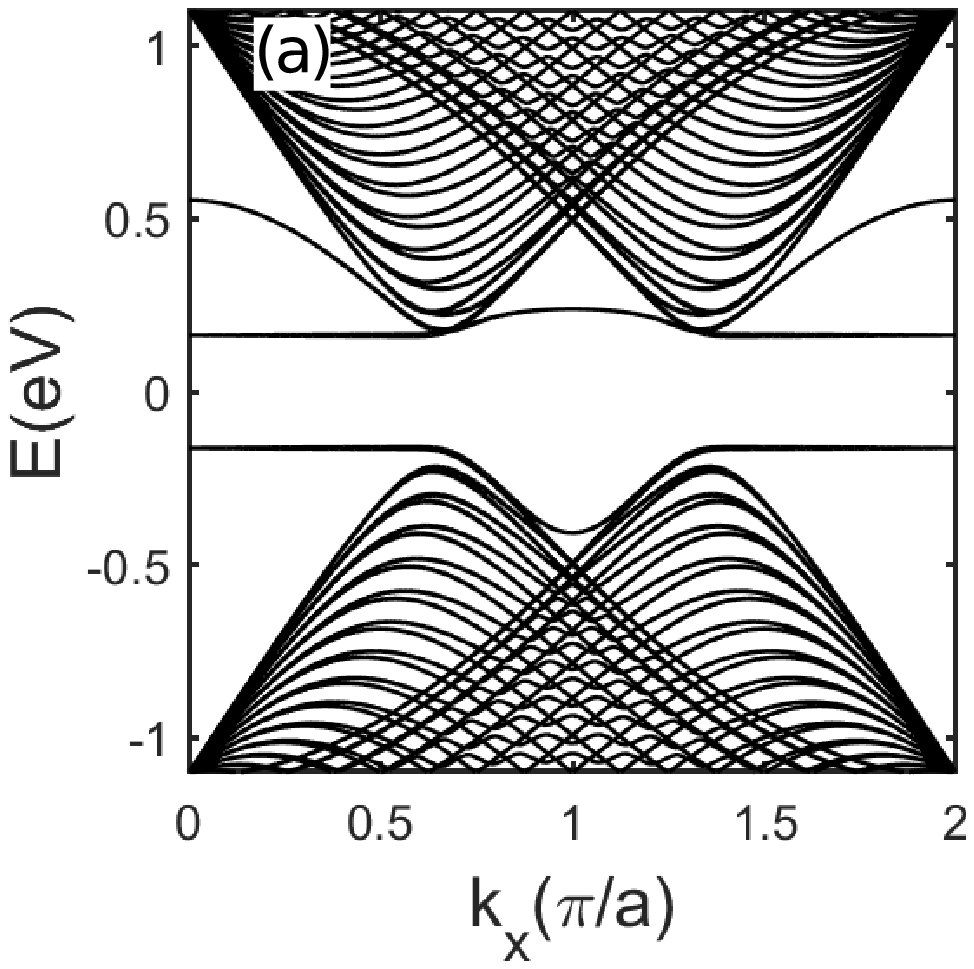}%
	\includegraphics[scale=0.27]{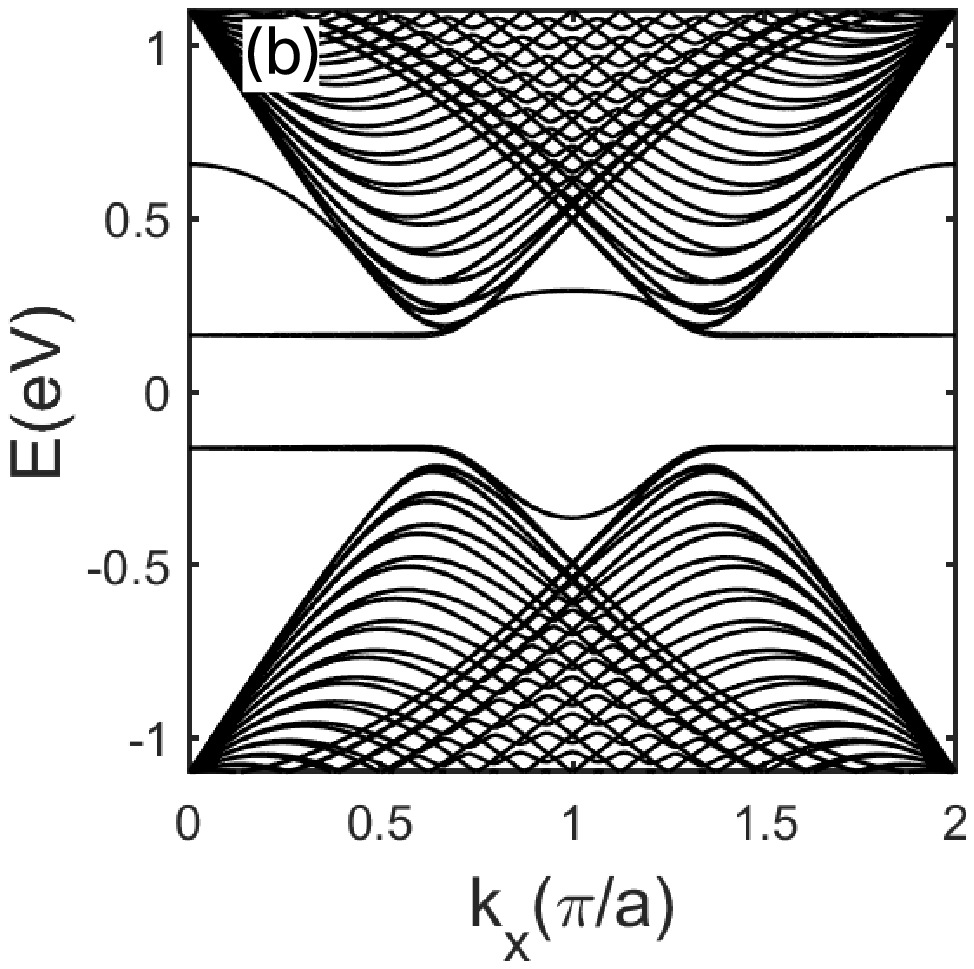}%
	\includegraphics[scale=0.27]{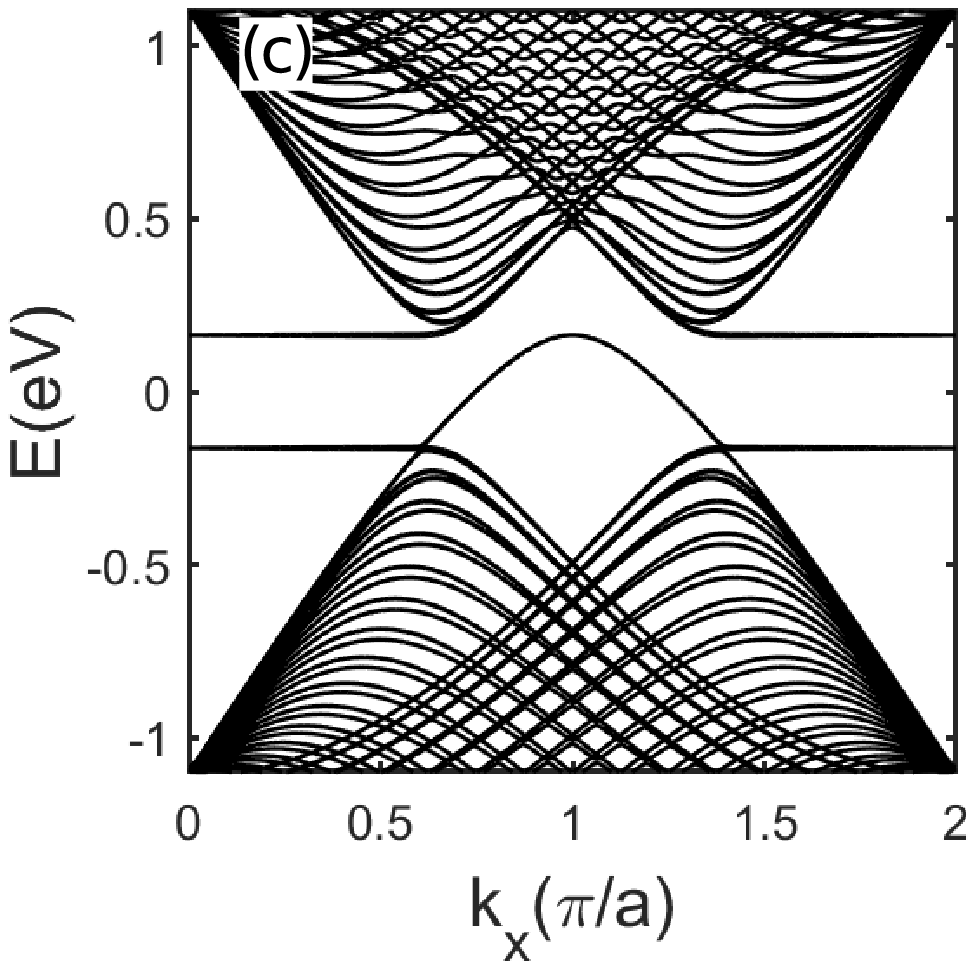}%
	\caption{Band spectra are plotted as a function of the wave vector $k_x$ for
	different on-site energies $E_{d}$ of line defect. (a)-(c) associates with case 2 with $E_{d}=0,~0.3t$ and $3t$, respectively. The other parameters are chosen as $\Omega=-0.15t$, $t_{so}=3.9\mathrm{meV}$, $N_{x}=121.80\mathrm{nm}$ and $N_{y}=17.39\mathrm{nm}$.}
\label{fig:band3}
	\end{figure}
	
Furthermore, when the sublattice potential is changed from $\Omega=0.15t$ to $\Omega=-0.15t$ by changing the direction of the electric field in case 2, variation trends of the band spectra are different from those of the system with $\Omega=0.15t$. In Fig.~\ref{fig:band3}(a), when $E_{d}=0$, the gapless state disappears and a band gap is opened, which is significantly different from the result shown in Fig.~\ref{fig:band2}(d). Obviously, there exists a transition from gapless state to band insulator for the silicene nanoribbon. Interestingly, this transition arises from the effect of the electric field due to the change of its direction, instead of the effect of the on-site energy of the line defect. When the on-site energy increases to $E_{d}=0.3t$, the band gap keeps unchanged. But with the on-site energy further increasing, for example $E_{d}=3t$ as shown in Fig.~\ref{fig:band3}(c), one flat band gradually bends upwards changing from downwards, thus this bending band fills in the band gap and induces a gapless state. Clearly, the electric field and the on-site energy have different effects on the band structure, which can be utilized to modulate the energy band from gapless states to band insulators.

Seen from Figs.~\ref{fig:band2}(a)-(c), for case 1 with the parameter $\Omega=0.15t$, we can see that the on-site energy of line defect doesn't induce a gapless state. Correspondingly, when the direction of the electric field is changed to opposite direction, namely $\Omega=-0.15t$, there still exists no gapless state. That is to say, the line defect configuration of case 1 can not be utilized to modulate the energy band from gapless states to band insulators.
	
\begin{figure}[t]
	\centering
	\includegraphics[scale=0.42]{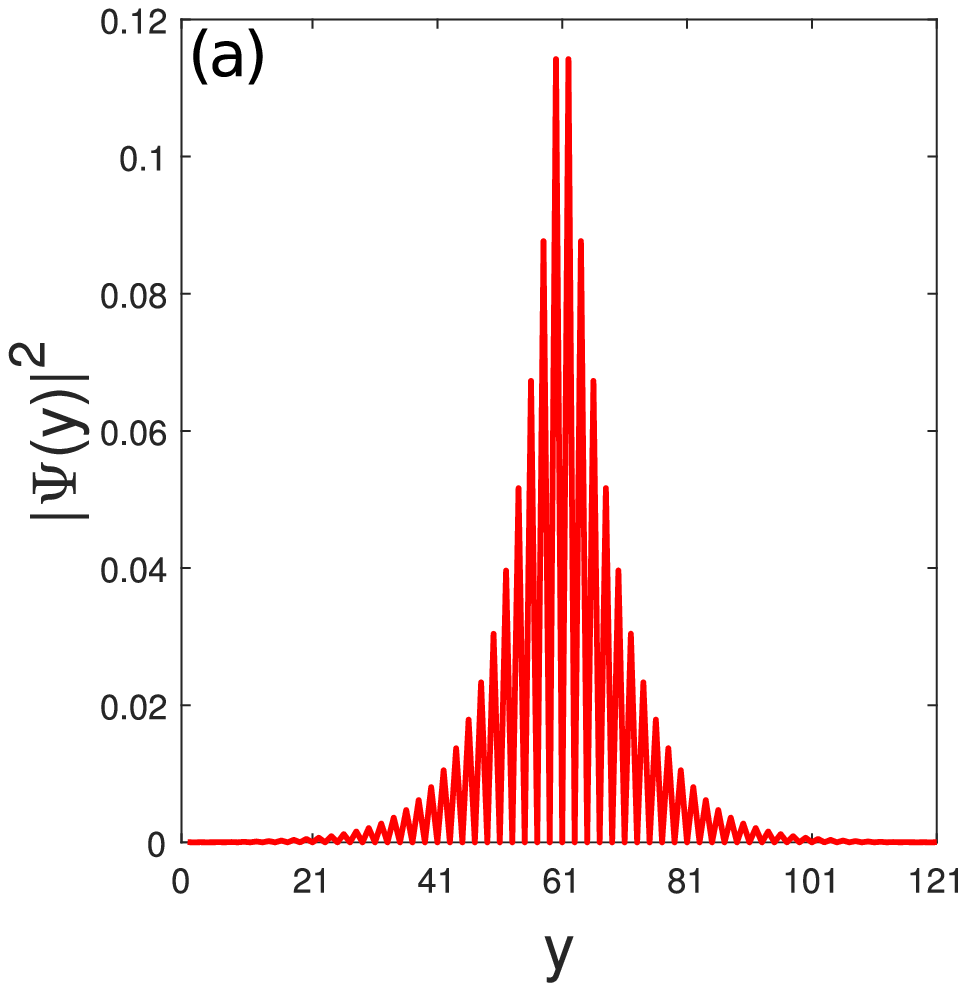}%
	\includegraphics[scale=0.42]{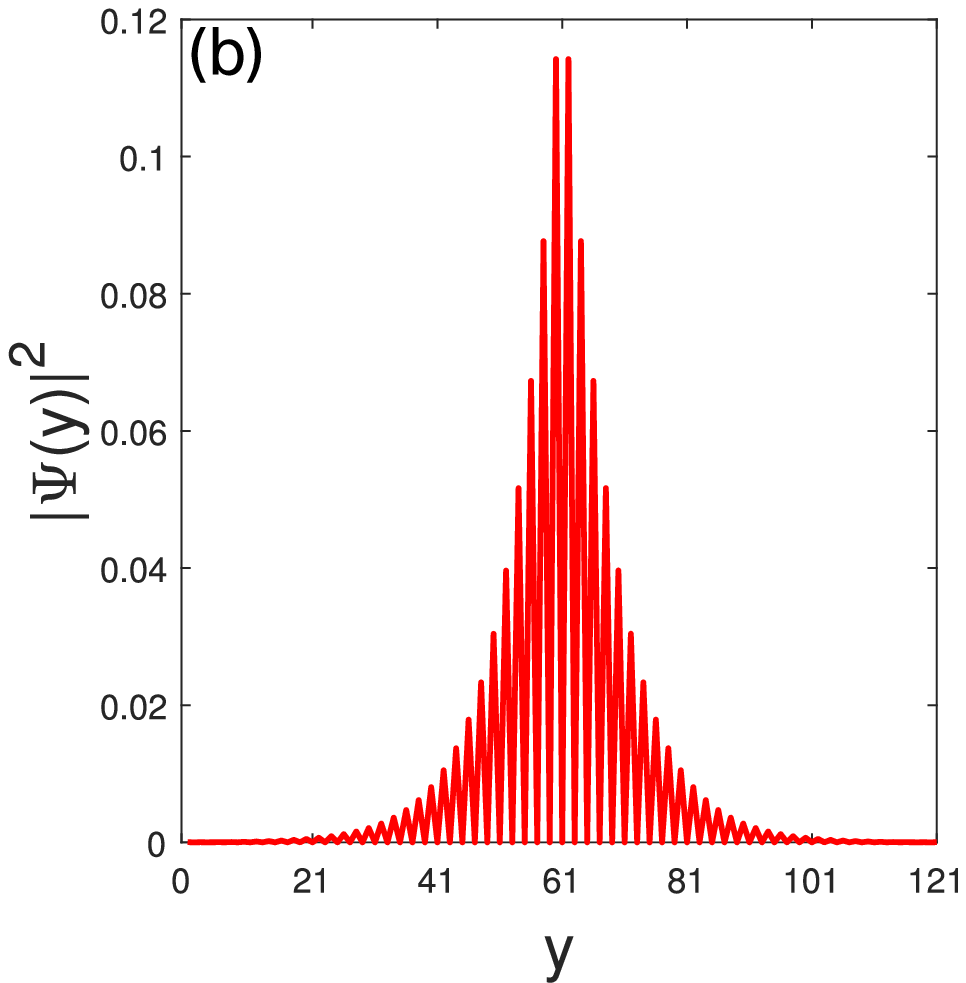}%
	\caption{The distribution probability of case 2 in real space is plotted as a function of the transverse coordinate $y$
for two different wave vectors (a) $k_{x}=0.62~\pi/a$ and (b) $k_{x}=1.38~\pi/a$. The other parameters are $\Omega=0.15t$, $E=-0.13t$, $E_{d}=0$, $N_{x}=121.80\mathrm{nm}$ and $N_{y}=20.07\mathrm{nm}$.}
\label{fig:wave1}
\end{figure}

As is well known, an effective method to investigate the gapless state is to examine the distribution probability in real space. In Fig.~\ref{fig:wave1}, we plot the distribution probability as a function of the transverse coordinate $y$ for two different wave vectors. Here, the Fermi energy is fixed as $E =-0.13t$, which corresponds to the energy level locating inside the band gap and intersecting with the gapless state at two momenta of $k_{x} = 0.62~\pi/a $ and $1.38~\pi/a$. Note that the positions of the bottom-edge and top-edge sites are presented by $y=1$ and $y=121$, respectively, and the position of the line defect is at $y=61$. 	
	
In Fig.~\ref{fig:wave1}(a), when $k_{x}$ is $ 0.62~\pi/a$, it can be observed that the probability mainly localizes around
the center of the nanoribbon near the line defect when the Fermi energy is close to the top valence band.
Especially, the probability density of the line defect, namely the 61st site, is zero. Therefore, the gapless state mainly arises from
the contribution of the neighbor sites of the line defect. Furthermore, the same distribution probability of the wave functions
can be found for the wave vector $k_{x}=1.38~\pi/a$ in Fig.~\ref{fig:wave1}(b). 	
Based on the above analysis, we can use the metallicity of line defects to make interesting silicene-based nano devices, and effectively modulate the transport properties of the devices by adjusting the gate voltage under line defects. Since silicene is easily compatible with silicon-based semiconductor devices, the proposed silicene-based devices with line defects may have a wide range of potential applications in the field of nanoelectronics devices in future. Thus, line defects can have significant contributions in the application of silicene-based devices.
	
\subsection{Band spectra for different spin-orbit couplings}
\begin{figure}[t]
	\centering
	\includegraphics[scale=0.4]{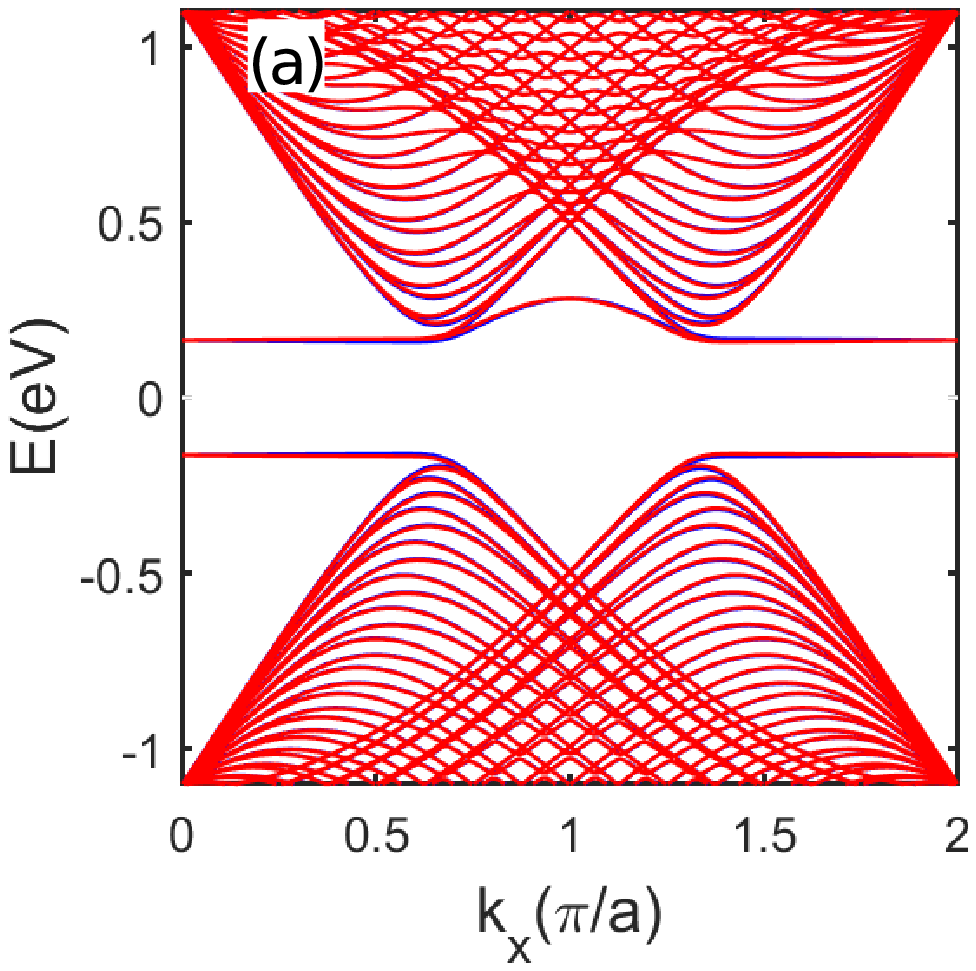}%
	\includegraphics[scale=0.4]{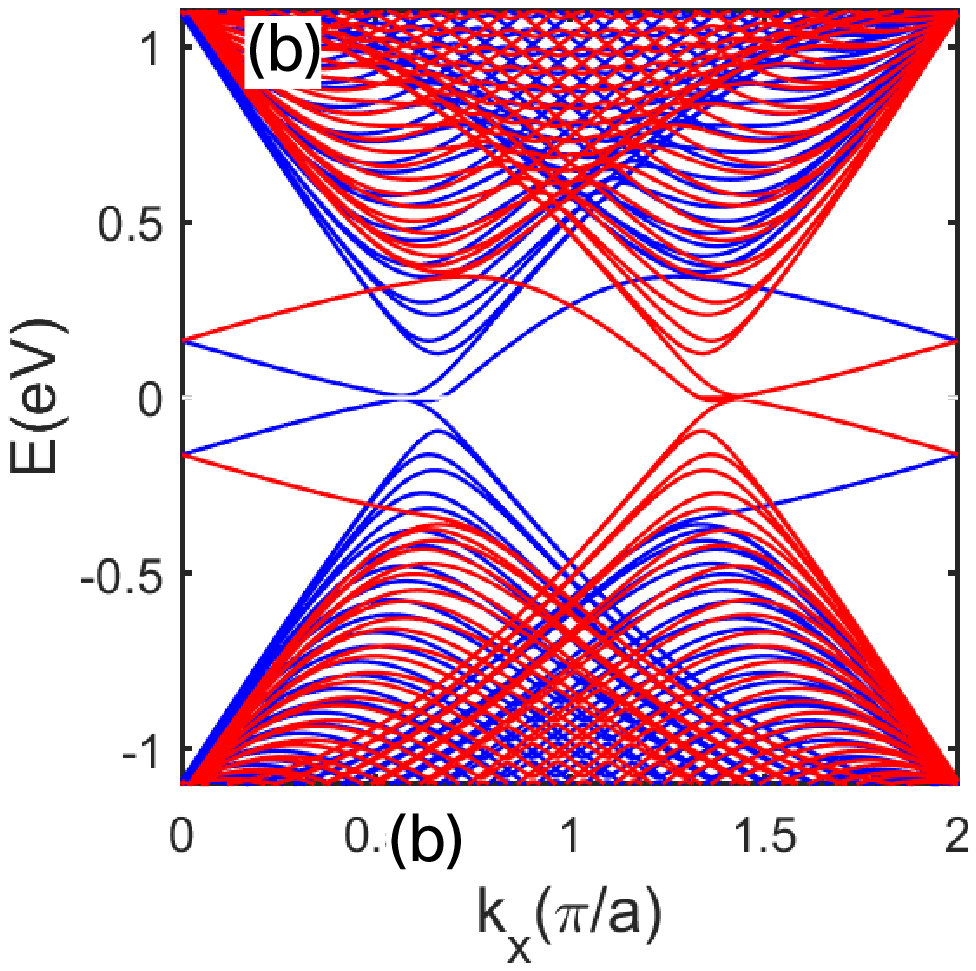}%

	\caption{Band spectra of case 2 are plotted as a function of the wave vector $k_x$ for different amplitudes of the spin-orbit coupling (a) $t_{so}=3.9\mathrm{meV}$, (b) $t_{so}=180\mathrm{meV}$. The other parameters are chosen as $E_{d}=3t$, $\Omega=0.15t$, $N_{x}=121.80\mathrm{nm}$ and $N_{y}=17.39\mathrm{nm}$. Blue and red lines refer to spin-up and spin-down energy bands, respectively.}
\label{fig:band4}
	\end{figure}
	
It is proposed that the exchange field and electric field applied in silicene systems can induce phase transition~\cite{ Shakouri2015}, which is similar to the effect of the spin-orbit coupling (SOC). A fundamental question is that whether the gapless state can be generated by SOC~\cite{Xian_2017, Zhiming2015c}. In addition, the research on the effect of SOC is meaningful for investigation of other low warpage honeycomb structures with SOC and line defects,
like germanene\cite{Liu2011,Tabert2013aa}.
	
\begin{figure}[t]
	\centering
	\includegraphics[scale=0.4]{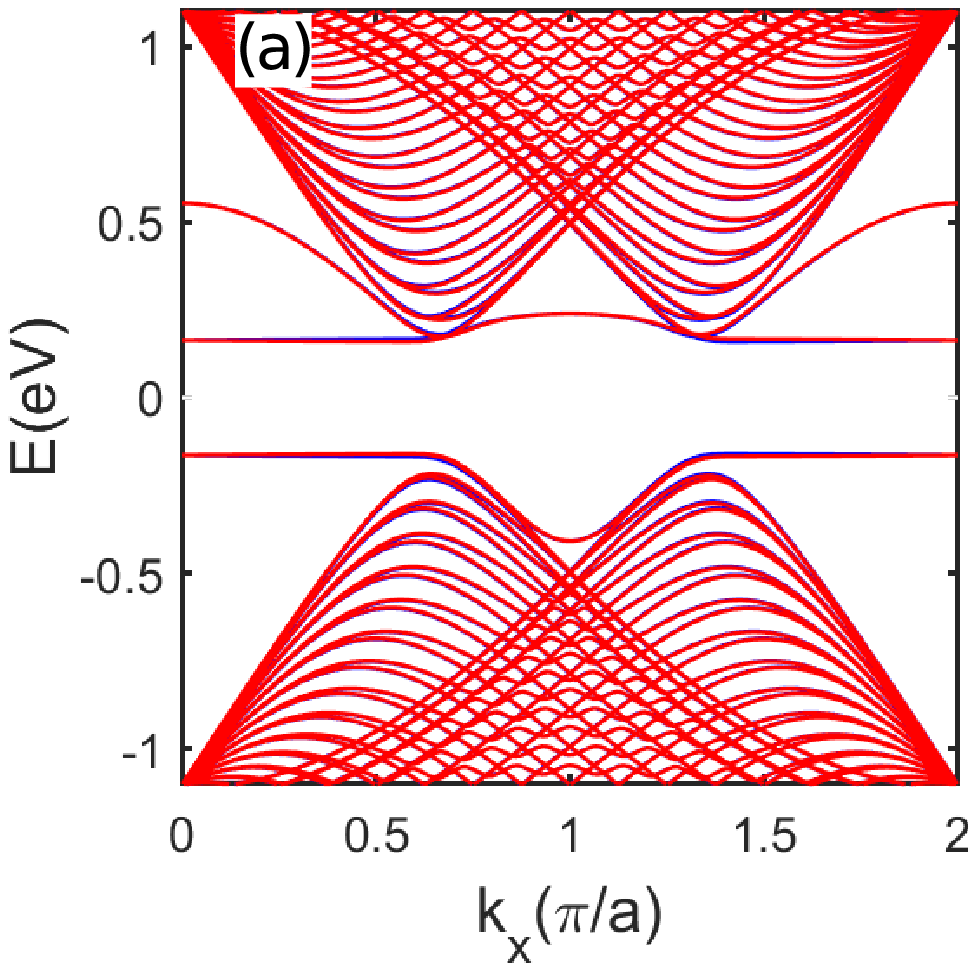}%
	\includegraphics[scale=0.4]{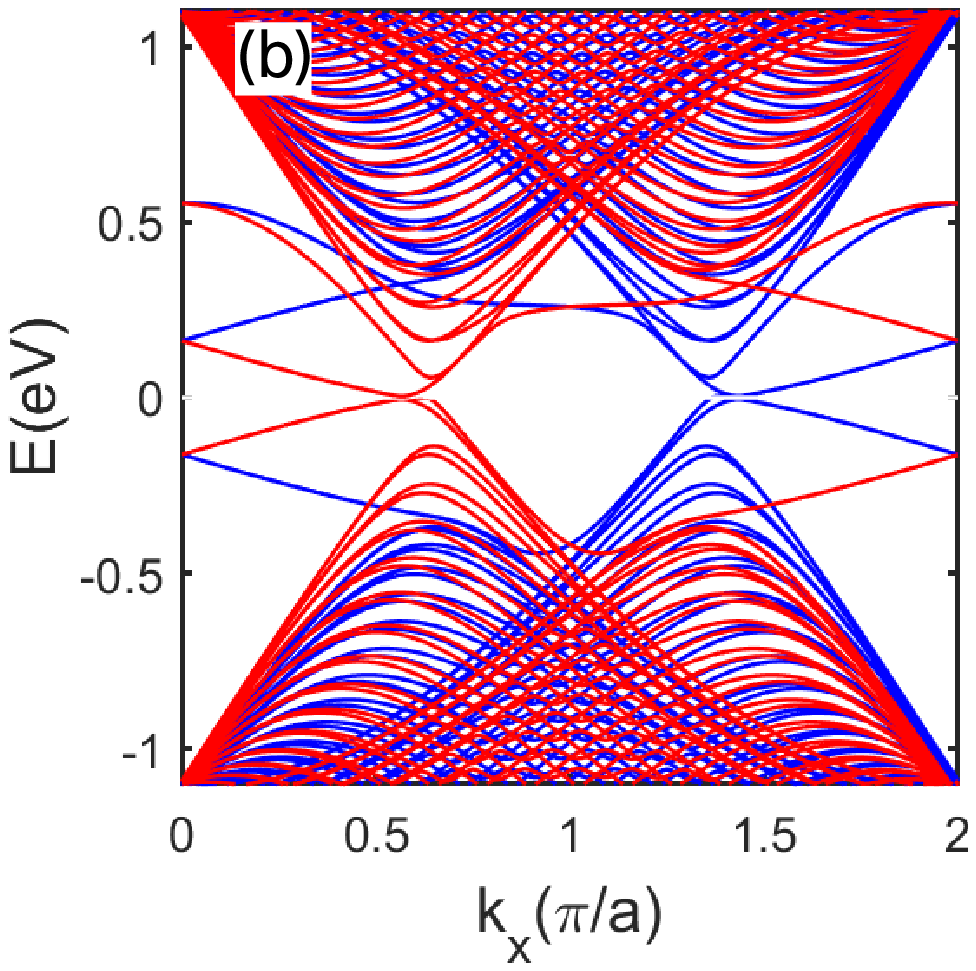}%
	\caption{Band spectra of case 2 are plotted as a function of the wave vector $k_x$ for different amplitudes of the spin-orbit coupling (a) $t_{so}=3.9\mathrm{meV}$, (b) $t_{so}=180\mathrm{meV}$. The other parameters are chosen as $E_{d}=0$, $\Omega=-0.15t$, $N_{x}=121.80\mathrm{nm}$ and $N_{y}=17.39\mathrm{nm}$. Blue and red lines refer to spin-up and spin-down energy bands, respectively.}
\label{fig:band5}
	\end{figure}

Motivated by this idea, we study the behavior of the band spectrum when the SOC of the silicene is tuned. Here, we only consider the defect configuration of case 2 with the width $N_{y} = 17.39\mathrm{nm}$. In order to better understand the effect of SOC, we plot the spin-dependent band of silicene with line defects, as shown in Figs.~\ref{fig:band4} and~\ref{fig:band5}. Note that blue and red lines refer to spin-up and spin-down energy bands, respectively. In Fig.~\ref{fig:band4}, $E_{d}=3t$ and $\Omega=0.15t$, with SOC increasing from $ t_{so}=3.9\mathrm{meV}$ to $ 180\mathrm{meV}$,  the energy bands gradually separate into spin-dependent bands. What's more, spin-up energy bands occupy the band gap of K valley, and spin-down energy bands occupy the band gap of K' valley, which form a gapless state. Obviously, the silicene system gradually evolves into a spin-valley-polarized metal (SVPM) from a band insulator due to the effect of SOC.
Furthermore, similar behaviors of band spectra can be found in Fig.~\ref{fig:band5}. When SOC increases from $ t_{so}=3.9\mathrm{meV}$ to $180\mathrm{meV}$, a spin-valley-polarized metal phase also occurs. Differently, the spin polarization of valleys becomes opposite, which attributes to
the effect of the electric field with the parameter $\Omega=-0.15t$.
	
Based on the analysis above, we can make transport channel switch by utilizing SOC, the exchange field and electric field. It can change the physical properties of quantum nano channels in the case of mirror symmetry. The phase transition is very interesting, which can provide us with some methods to make electric channel switch and have a guiding significance for the study of transport properties of the warped honeycomb structures with line defects.
	
\subsection{The conductance for different on-site energies}
\begin{figure}[t]
	\centering
	\includegraphics[scale=0.38]{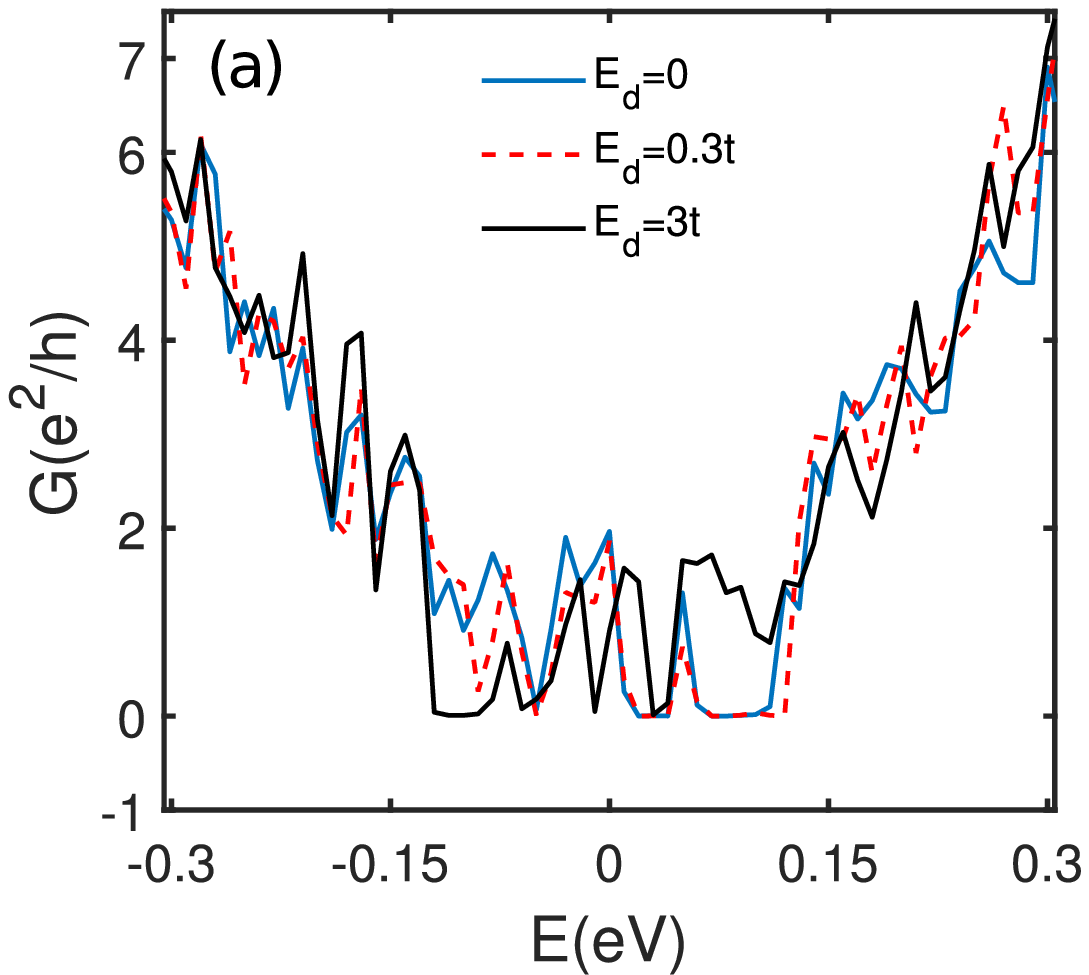}%
	\includegraphics[scale=0.38]{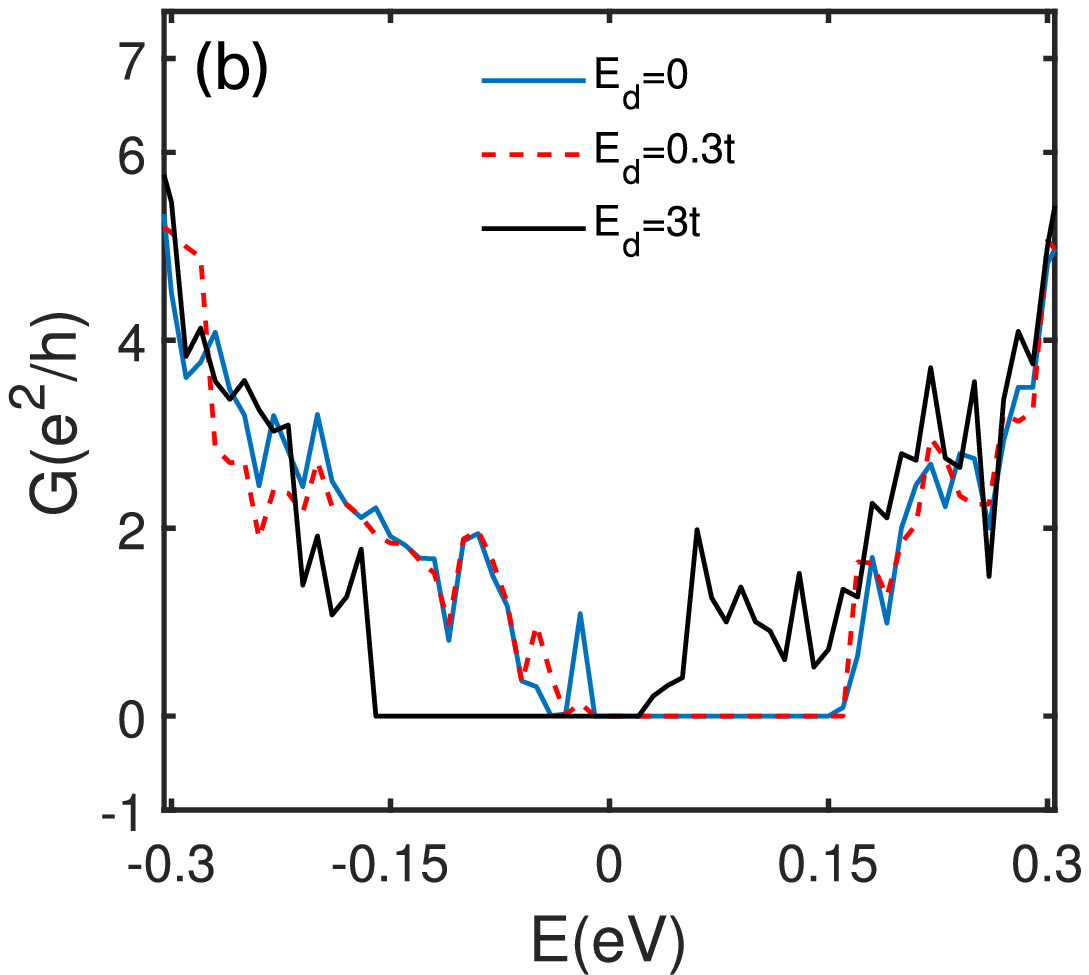}%
		
	\includegraphics[scale=0.38]{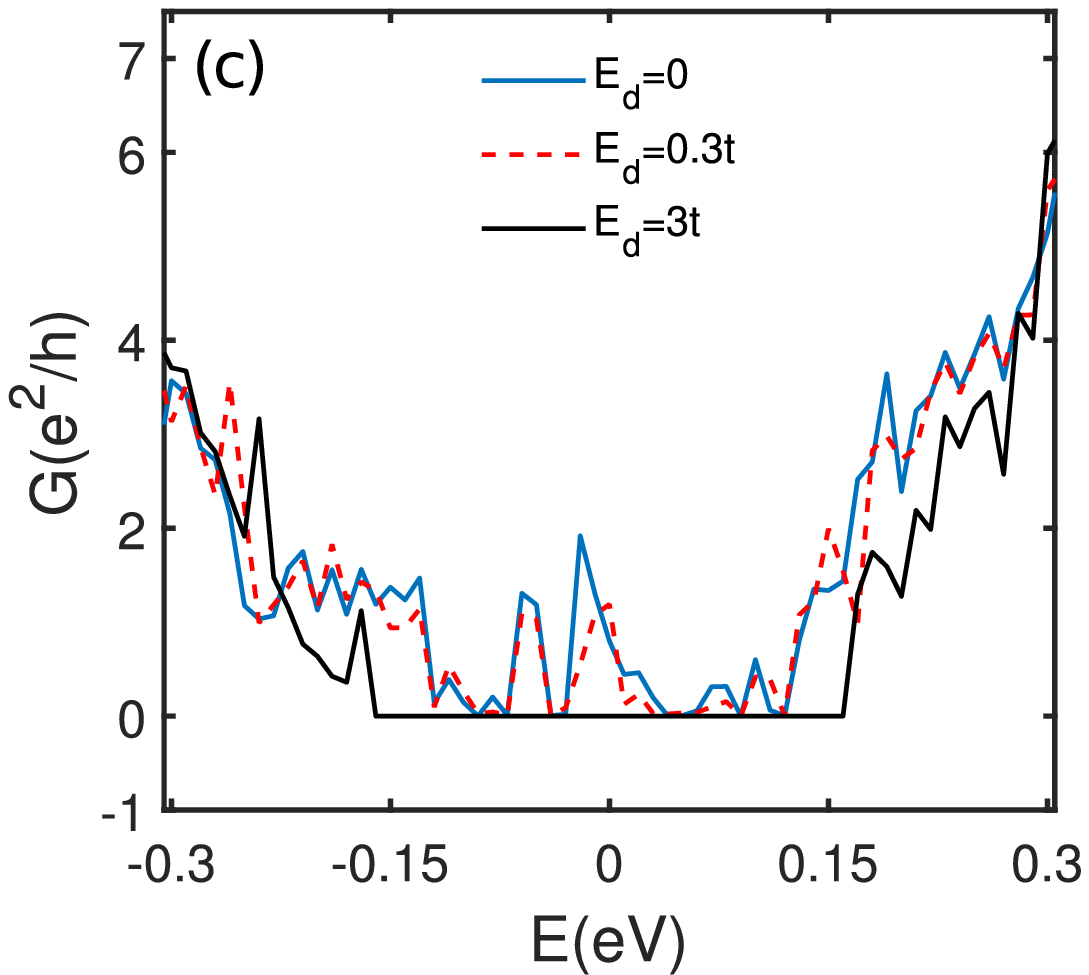}%
	\includegraphics[scale=0.38]{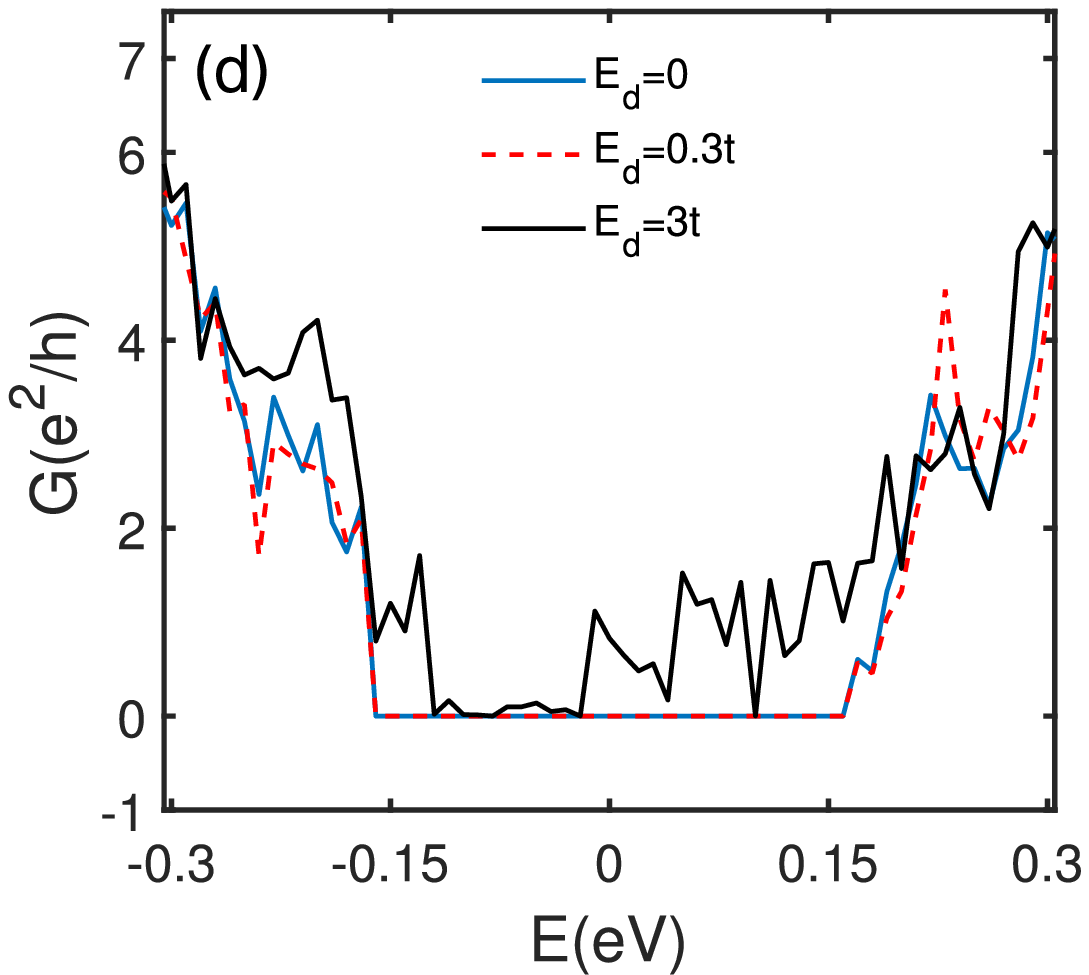}%
	\caption{The conductance $G$ is plotted as a function of Fermi energy for various on-site energies for a line defect. (a)-(b) associates with case 1 with $\Omega=0$ and $\Omega= 0.15t$, respectively. (c)-(d) associates with case 2 with $\Omega=0.15t$ and $-0.15t$, respectively. The other parameters are chosen as $t_{so}=3.9\mathrm{meV}$, $N_{x}=121.80\mathrm{nm}$ and $N_{y}=17.39\mathrm{nm}$.}
\label{fig:trans2}
\end{figure}

In Fig.~\ref{fig:trans2}, we plot the conductance as a function of Fermi energy
with different staggered potentials and on-site energies for
the line defect with two configurations. For case 1 with $ \Omega=0$ in Fig.~\ref{fig:trans2}(a), an asymmetrical conductance
is observed near the Dirac point $E=0$, which indicates that the particle-hole symmetry is broken due to the existence
of the line defect. Within the energy regime from $E=-0.17~\mathrm{eV}$ to $0.17~\mathrm{eV}$, the conductance fluctuates and the maximum value will not exceed $2G_{0}$ for different on-site energies of the line defect. With further increasing of the Fermi energy, the conductance gradually increases and shows an oscillating behavior. When the  staggered sublattice potential is $\Omega = 0.15t$, the conductance keeps zero within the energy regime $-0.02<E<0.17~\mathrm{eV}$ for the case $E_d=0$ and $E_d=0.3 t$ (see Fig.~\ref{fig:trans2}(b)). When $E<-0.02~\mathrm{eV}$ or $E>0.17~\mathrm{eV}$, the conductance gradually increases with increasing of the amplitude of the Fermi energy. While the on-site energy is increased to $E_d=3t$, the variation trend of the conductance becomes opposite. When $-0.17<E<0.02~\mathrm{eV}$, the conductance keeps zero. When $E<-0.17~\mathrm{eV}$ or $E>0.02~\mathrm{eV}$, the conductance also gradually increases with increasing of the amplitude of the Fermi energy. The opposite transport behaviors can be understood from the band structures shown in Figs.~\ref{fig:band2}(a)-(c). We can see that there exists a band gap within the energy regime $-0.02<E<0.17~\mathrm{eV}$ for the case $E_d=0$ and $E_d=0.3 t$, which associates with the zero value of the conductance in Fig.~\ref{fig:trans2}(b). While the on-site energy makes the energy regime of the band gap change to $-0.17<E<0.02~\mathrm{eV}$ when $E_d=3t$, which gives opposite transport behaviors. It means that the on-site energies can be utilized to modulate the transport property of the silicene nanoribbon with a line defect. However, when the staggered potential is changed from $\Omega = 0.15t$ to $\Omega =-0.15t$, it can not induce new band gap for the defect configuration of case 1.

For the defect configuration of case 2, we can see the combined effect of the staggered potential and the on-site energies on the conductance. When $E_d=0$
and $E_d=0.3t$ for the condition $\Omega = 0.15t$ (see Fig.~\ref{fig:trans2}(c)), the conductance is not zero within the energy regime $-0.17<E<0.17~\mathrm{eV}$ due to the existence of the gapless state between the band gap shown in Figs.~\ref{fig:band2}(d)-(f). When the on-site energy increase to $E_d=3t$, the conductance significantly changes to zero within the energy regime $-0.17<E<0.17~\mathrm{eV}$ due to the large band gap.
When the staggered potential is changed from $\Omega = 0.15t$ to $\Omega =-0.15t$, the conductance decreases to zero for the conditions $E_d=0$
and $E_d=0.3t$ when $-0.17<E<0.17~\mathrm{eV}$ because of the existence of the band gap, as shown in Figs.~\ref{fig:band3}(a)-(b). Interestingly, the gapless state induced by the on-site energy $ E_d=3t$ can result in a nonzero conductance when $-0.17<E<0.17~\mathrm{eV}$. It means that the transport property of the silicene nanoribbon can be effectively modulated by changing the staggered potential and on-site energies.

\begin{figure}[t]
	\centering
	\includegraphics[scale=0.38]{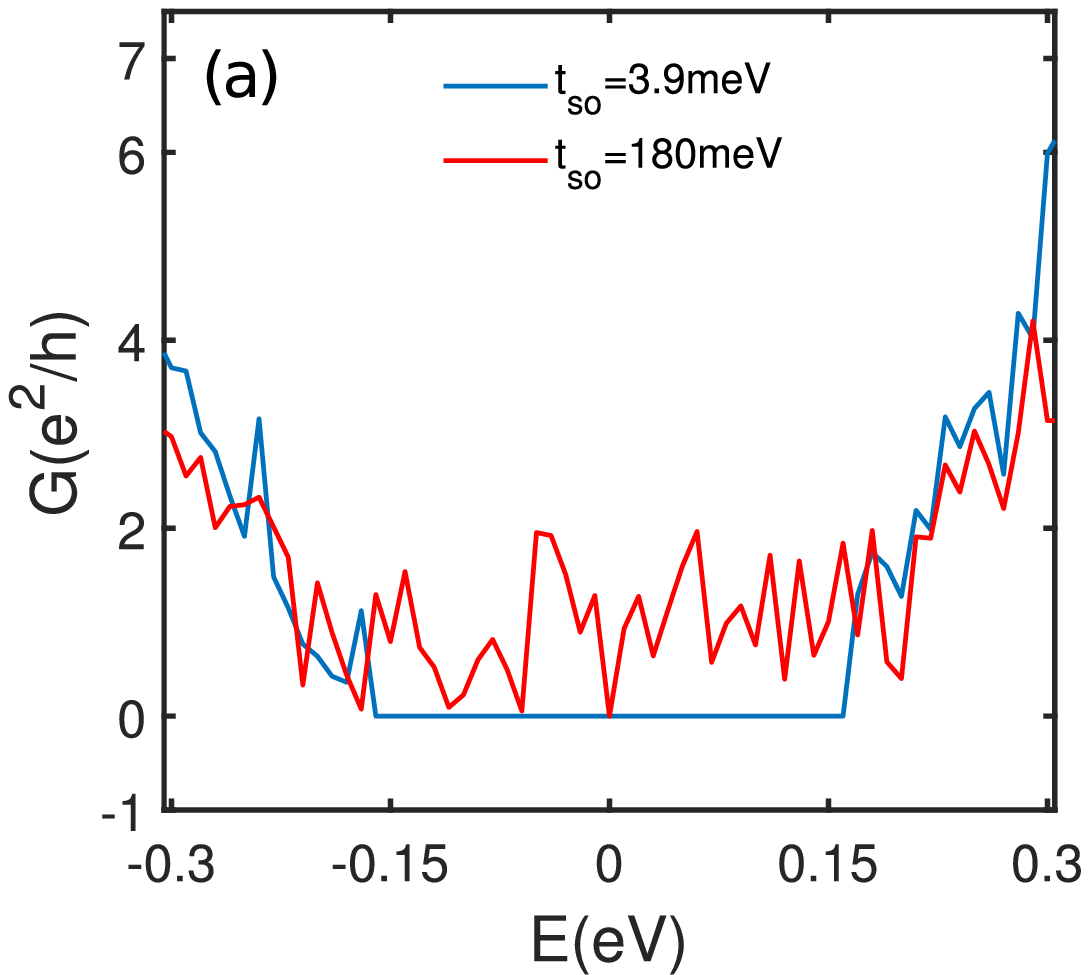}%
	\includegraphics[scale=0.38]{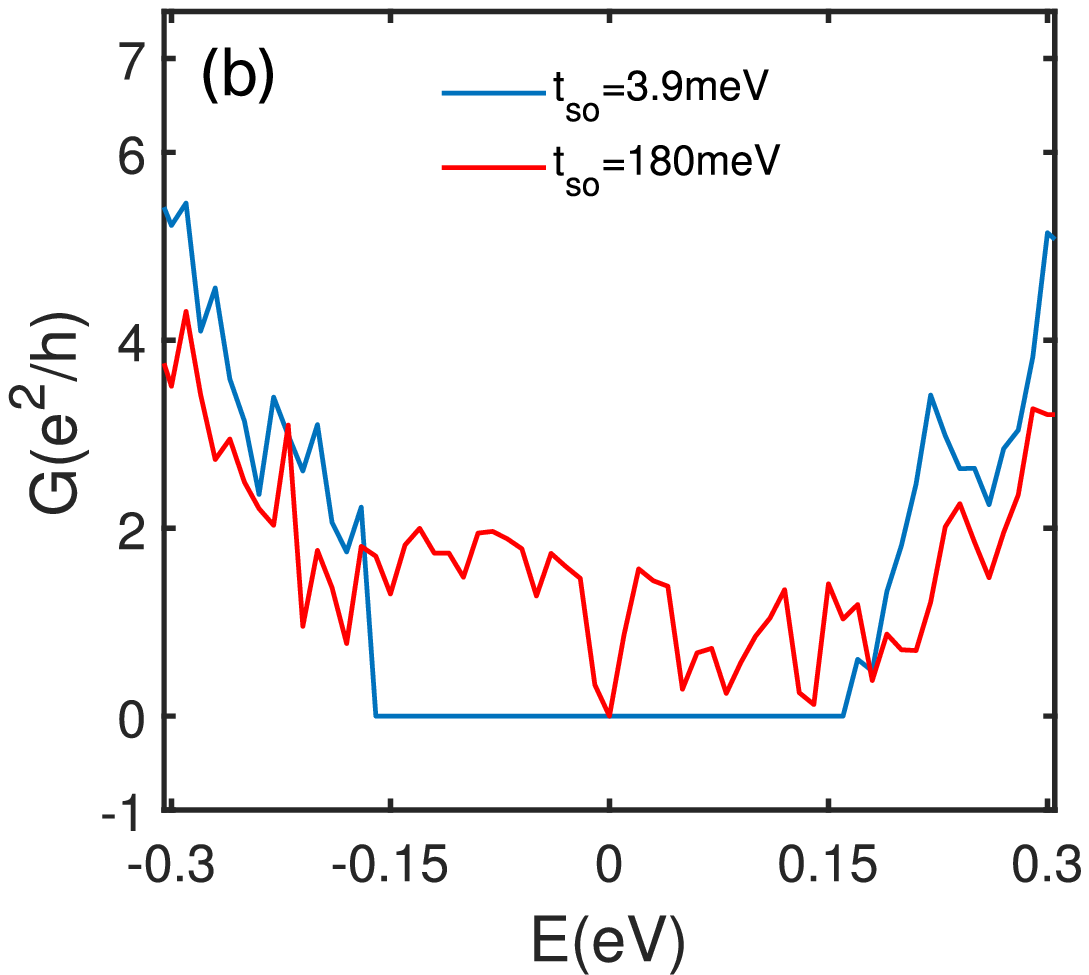}%
	\caption{The conductance $G$ of case 2 is plotted as a function of Fermi energy for different spin-orbit couplings (a) $E_d=3t$, $\Omega= 0.15t$ and (b) $E_d=0$, $\Omega=-0.15t$. The other parameters are chosen as $N_{x}=121.80\mathrm{nm}$ and $N_{y}=17.39\mathrm{nm}$.}
\label{fig:trans3}
\end{figure}

Next, we consider the influence of SOC on the transport properties for the defect configuration of case 2, as shown in Fig.~\ref{fig:trans3}.
We can see that the conductance is zero within the energy regime $-0.17<E<0.17~\mathrm{eV}$ because of the existence of the band gap when $t_{so}=3.9\mathrm{meV}$ (see Fig.~\ref{fig:band4}(a)). With SOC increasing to $180\mathrm{meV}$, the spin-dependent energy bands occupy the band gap
and form a gapless state, which result in the nonzero conductance in the same energy regime. Similarly, when the parameters are $E_d=0$, $\Omega=-0.15t$,
zero conductance can also be modulated by increasing the amplitude of SOC. Obviously, when the staggered potential and on-site energies are fixed,
the spin-orbit coupling can be utilizing to modulate the conductance of the silicene nanoribbon with a line defect.
Thus, an electric channel can be constructed by using the line defect, and importantly it can be turned on or off easily by changing the on-site energy, the staggered potential and SOC.

\section{SUMMARY}
We study the band structures and the transport property of the silicene nanoribbon with a line defect, taking into account the sublattice symmetry breaking.
The flat band bends downwards or upwards due to the effect of the line defect. The spin-orbit coupling induces the quantum spin Hall states. Especially, the energy band depends on the distance between the line defect and the edge of the nanoribbon. The conductance has a maximum value when the line defect is located at the middle position of the silicene nanoribbon. The effects of the on-site energies on the band spectra of the two defect configurations are different. For the defect configuration of case 1, there always exists one band gap, the middle flat subband bends downwards first and then upwards with increasing of the amplitude of the on-site energy. While for the defect configuration of case 2, there exists a gapless state, which can be broken and form a band gap due to the effect of the on-site energy.
At this case, spin-dependent gapless subbands can occupy the band gap by increasing the amplitude of the spin-orbit coupling. When the sublattice potential is changed from positive to negative values, the gapless subband disappears and a band gap is opened. A valence subband bends upwards and forms a gapless state with increasing of the on-site energy, which is different from the case of the sublattice potential with positive values. If the on-site energy is zero, the band gap can also be occupied by the spin-dependent gapless subbands due to the effect of the spin-orbit coupling. The variation trends of the conductance including zero conductance can be well understood in terms of the combined effect of the sublattice potential, the on-site energy and spin-orbit couplings on the band structures. It means that one can effectively modulate the transport property of the silicene nanoribbon with a line defect by utilizing the sublattice potential, the on-site energy and spin-orbit couplings.

\begin{acknowledgments}
This work was supported by National Natural Science Foundation of China (Grant No. 11574067).
\end{acknowledgments}

\end{document}